\documentclass[journal]{IEEEtran}
\usepackage{cite}
\usepackage{amsmath,amssymb,amsfonts}
\usepackage{algorithm}
\usepackage{algpseudocode}
\usepackage{graphicx}
\usepackage{textcomp}
\usepackage{subfigure}
\usepackage{epstopdf}
\usepackage{tablefootnote}
\usepackage[spaces,hyphens]{url}
\usepackage[colorlinks,allcolors=blue]{hyperref}
\usepackage[table]{xcolor}
\usepackage{booktabs}
\usepackage{tablefootnote}
\usepackage{marvosym} % para usar \Cross
\everymath{\displaystyle}
\def\BibTeX{{\rm B\kern-.05em{\sc i\kern-.025em b}\kern-.08em
    T\kern-.1667em\lower.7ex\hbox{E}\kern-.125emX}}
\begin{document}
\title{Discontinuous Reception with Adjustable Inactivity Timer for IIoT}
\author{{David E. Ruíz-Guirola, \IEEEmembership{Graduated Student Member, IEEE}, Carlos A. Rodríguez-López, Onel L. A. López, \IEEEmembership{Senior Member, IEEE}, Samuel Montejo-Sánchez, \IEEEmembership{Senior Member, IEEE}, Vitalio Alfonso Reguera, \IEEEmembership{Senior Member, IEEE}
and Matti Latva-aho, \IEEEmembership{Senior Member, IEEE}
}
\thanks{David E. Ruíz-Guirola, Onel L. A. López and Matti Latva-aho are with the Centre for Wireless Communications University of Oulu, Finland. \{David.RuizGuirola, Onel.AlcarazLopez, Matti.Latva-aho\}@oulu.fi.  
Samuel Montejo-Sánchez is with the {Instituto Universitario de Investigación y Desarrollo Tecnológico, Universidad Tecnológica Metropolitana, Santiago, Chile}.\{smontejo@utem.cl\}.  Carlos A. Rodríguez-López is with the {Department of Electronics and Telecommunications, Universidad Central ``Marta Abreu'' de Las Villas}, Cuba. \{crodrigz@uclv.edu.cu\}. Vitalio Alfonso Reguera is with the Department of Information Technologies, {Universidad Tecnologica del Uruguay (UTEC)}.  
\{vitalio.alfonso@utec.edu.uy\}.\\
\indent This work has been partially supported by the Finnish Foundation for Technology Promotion and the Research Council of Finland (former Academy of Finland) 6G Flagship Programme (Grant Number: 346208), and in Chile by ANID FONDECYT Regular No.1241977.
}
}

%This work has been partially supported in Chile by ANID FONDECYT Iniciación No. 11200659 and FONDEQUIP-EQM180180, %and Collaborative Research Activities between PIDi/UTEM and FIE/UCLV, 
%in Brazil by CNPq (402378/2021-0, 305021/2021-4), Print CAPES-UFSC “Automation 4.0”, and RNP/MCTIC (Grant 01245.010604/2020-14), and in Finland by 6Genesis Flagship (Grant no. 318927) and Tekniikan Edistämissäätiön.}}		
%\thanks{This paragraph of the first footnote will contain the date on 
%which you submitted your paper for review. It will also contain support 
%information, including sponsor and financial support acknowledgment. For 
%example, ``This work was supported in part by the U.S. Department of 
%Commerce under Grant BS123456.'' }
%\thanks{The next few paragraphs should contain 
%the authors' current affiliations, including current address and e-mail. For 
%example, F. A. Author is with the National Institute of Standards and 
%Technology, Boulder, CO 80305 USA (e-mail: author@boulder.nist.gov). }
%\thanks{S. B. Author, Jr., was with Rice University, Houston, TX 77005 USA. He is 
%now with the Department of Physics, Colorado State University, Fort Collins, 
%CO 80523 USA (e-mail: author@lamar.colostate.edu).}
%\thanks{T. C. Author is with 
%the Electrical Engineering Department, University of Colorado, Boulder, CO 
%80309 USA, on leave from the National Research Institute for Metals, 
%Tsukuba, Japan (e-mail: author@nrim.go.jp).}

\maketitle

\begin{abstract}
Discontinuous reception (DRX) is a key technology for reducing the energy consumption of industrial Internet of Things (IIoT) devices. Specifically, DRX allows the devices to operate in a low-power mode when no data reception is scheduled, and its effectiveness depends on the proper configuration of the DRX parameters. In this paper, we characterize the DRX process departing from a semi-Markov chain modeling and detail two ways to set DRX parameters to minimize the device power consumption while meeting a mean delay constraint. The first method exhaustively searches for the optimal configuration, while the second method uses a low-complexity metaheuristic to find a sub-optimal configuration, thus considering ideal and practical DRX configurations. Notably, within the DRX parameters, the inactivity timer (IT) is a caution time that specifies how long a device remain active after the last information exchange. Traditionally, a device implementing DRX will restart the IT after each data reception as a precedent to a low-power mode. The usual approach lies in restarting the IT whenever new data is received during this cautious period, which might sometimes needlessly extend the active time. Herein, we propose a more efficient method in which the transmit base station (BS) explicitly indicates restarting the timer through the control channel only when appropriate. The decision is taken based on the BS's knowledge about its buffer status. We consider Poisson and bursty traffic models, which are typical in IIoT setups, and verify the suitability of our proposal for reducing the energy consumption of the devices without significantly compromising the communication latency through extensive numerical simulations. Specifically, energy saving gains up to 30\% can be obtained regardless of the arrivals rate and delay constraints. 
\end{abstract}


\section{Introduction}
\label{sec:introduction}

%\IEEEPARstart{A}{dopting} autonomous wireless devices has evolved as a major trend for the new called industrial Internet of things (IIoT).  
\IEEEPARstart{T}{he} new industrial Internet of Things (IIoT) %or Industry 4.0 
refers to interconnected devices networked together with computers' industrial applications, including manufacturing and energy management~\cite{kong2022deep}. 
IIoT is production-oriented and designed to reduce production costs and promote manufacturing efficiency enabling intelligent industrial operations% using advanced data analytics for transformational business outcomes, and it is redefining the landscape for business and individuals alike%, unlike Internet of Things (IoT), which is often human-centric services
~\cite{mao2021energy}. 
IIoT 
paradigm has evolved with the massive deployment of autonomous wireless devices, encouraged by significant improvements added by IoT in other sectors. 
Its widespread application %~\cite{kong2022deep}  
and rapid development %~\cite{xu2022energy}
%, with 
connecting %massive IIoT 
massive number of 
devices  
%IIoT technologies have attracted much attention from both industry and academics due to its widespread application and rapid development~\cite{xu2022energy,kong2022deep}, where 
generate tremendous amount of data and signals used for sensing, controlling, system maintenance, and data analysis~\cite{kong2022deep,xu2022energy,mao2021energy}. %, while connecting massive IIoT devices. 
%Meanwhile, supporting a huge amount of connected devices is indeed a challenging task, hence, in
Considering the limited battery of IIoT devices~\cite{khan2020industrial}, energy efficiency is a crucial problem in several innovative IIoT systems. 
%Especially, %it is a two-fold problem, on the one hand, 
{Indeed,} 
excessive energy consumption may shorten the devices' lifetime leading to system  malfunction%while sensing%, computing and communicating tasks executed by the devices can also lead to an increasing carbon footprint from the whole system’s perspective
~\cite{mao2021energy}, {thus proper mechanisms to reduce the energy consumption are mandatory}~\cite{xu2022energy}.  

In recent years, there have been many approaches to {reduce the energy consumption of IIoT devices, including}, the discontinuous reception (DRX) mechanism~\cite{eDRX, ruiz2021drx, aghdam2021traffic, moradi2021improving, sultania2021optimizing}. {A device supporting DRX can operate in two modes: idle and continuous reception. In idle mode, the device enters a ``deep sleep'' state when not receiving any data transmission requests at the physical downlink control channel (PDCCH), resulting in intermittent listening to conserve battery. However, if a packet arrives during deep sleep, the device must reconnect with the Base Station (BS) before transmitting the packet, incurring unnecessary delay. Configuring DRX is a challenging task as there is a trade-off between energy-saving and latency. A too-long DRX cycle can increase latency due to more packets arriving during sleep, while a too-short cycle reduces energy saving capabilities~\cite{8661498}.}

\subsection{Related works}

{Leveraging the DRX mechanism to prolong IIoT devices' battery lifetime has received considerable attention lately.} 
For instance, the authors 
in~\cite{ruiz2021drx} proposed a prediction-based configuration for DRX in human voice communications%, it is shown 
. They showed that %up to 30\% in 
energy savings of up to 30\% can be achieved by properly characterizing the data traffic pattern %beforehand. 
and configuring DRX accordingly. 
In~\cite{aghdam2021traffic}, the authors adaptively modified the DRX short and long sleep cycles %as well as the sleep time 
based on the average packet delay and incoming traffic, which allows achieving significant power
saving compared to the conventional DRX mechanism. 
In~\cite{moradi2021improving}, a %new flexible 
DRX 
%approaches is introduced, which is 
adjustable according to the data traffic and channel information was introduced to enable longer sleep opportunities at the devices. In addition, the authors in~\cite{sultania2021optimizing} presented a Markov chain model to evaluate the power consumption and latency of narrow band (NB)-IoT devices using power saving mode and DRX.  
The analytical model prediction {was shown to achieve} an average accuracy greater than 91\%. Meanwhile, they %is shown a useful tool to 
%automatically 
determined the %optimal 
parameter set for DRX in terms of latency and power consumption for various IoT use cases with different data traffic requirements. 
%All in all, knowing the traffic behaviour is been shown a helpful approach to reduce the power consumption.  
%The 
Notice that all the 
previous papers focused %their efforts 
on %trying to 
adaptively adjusting the DRX mechanism parameters such as the short cycle duration ($ T_s $), long cycle duration ($T_l$), and the short cycle timer ($ T_{sc} $) based on traffic statistics. %, while setting constraints, such as maximum allowed latency. 
All in all, exploiting the knowledge about data traffic behavior %has been shown to be 
is 
extremely useful for efficiently configuring DRX and obtaining important energy savings.

{In general, configuring DRX parameters in dynamic IIoT scenarios is challenging since the traffic patterns are unpredictable and diverse, and the timeliness and reliability performance requirements of such networks are especially stringent~\cite{sari2020industrial}. Note that efficiently exchanging information is typically critical to ensure correct and safe behavior of the controlled processes. Therefore, the communication network must be engineered to meet stringent delay deadlines, be robust to packet losses, and be safe and resilient to damages, especially in real-time applications~\cite{palattella2016internet}. However, DRX has not been optimized considering such specific requirements and characteristics of IIoT applications~\cite{khan2020industrial}. This calls for novel mechanisms balancing energy savings, responsiveness, and adaptability to dynamic conditions to optimize overall communication system performance.}

\subsection{Contributions}

%To the best of our knowledge, no prior research addresses the phenomenon of the IT reset explicitly from the perspective of the BS through the control channel. Herein, we aim to avoid keeping the device in a high consumption state for longer than necessary, thus, achieving greater energy savings. %by exploiting moments in which the device is idle. 
%Our main contributions are summarized as follows:
%\begin{itemize}
%    \item We introduce a DRX modeling based on a semi-Markov chain, while two types of data  traffic patterns are considered: (i) Poisson traffic and (ii) %a two-state Markov chain with 
%    bursty traffic. Mathematical derivations are also provided.  
%    \item We analyze the energy saving impact of controlling the IT reset process explicitly from the BS and propose a method for maximizing the energy saving by dynamically configuring the DRX parameters.
    %\item We propose a method for maximizing the energy saving by dynamically configuring the DRX parameters. %while the IT reset process is explicitly determined by the BS. %In addition, we study the impact in energy saving of controlling the IT reset process explicitly from the BS.
%    \item We evaluate the saving factor of the proposed method for the considered %MTC 
%    traffic patterns and delay constraints. In addition, the complexity of the proposed mechanisms %are analyzed in addition to the performance. 
%    is analyzed.
%\end{itemize}
%Ideally, 
Noteworthy, 
the BS can indicate the devices to %switch the parameters among these sets of configurations 
modify their DRX parameters.  
%in a dynamic way. 
However, %there are several issues with this approach. Most of the time the 
a key challenge lies in the fact that the data 
traffic is commonly heterogeneous, which makes it difficult to select the optimal DRX parameters. A more %realistic 
appealing 
way %is to consider adapting the configuration 
may lie in configuring the DRX mechanism 
from the Medium Access Control-Physical layer (MAC-PHY) point of view~\cite{notes}. %For a specific device, since 
In particular, the BS possesses knowledge of the downlink  buffer status for each device, enabling it to identify periods with data traffic and estimate the duration needed for transmitting buffered data. Notably, if the BS determines that the remaining active time is adequate for transmitting the buffered data, it is more efficient to refrain from resetting the inactivity timer (IT) to avoid unnecessary PDCCH-only slots without any grant~\cite{huawei20193gpp}.

\begin{table}
\caption{List of Symbols}
%\label{table1}
\setlength{\tabcolsep}{3pt}
\centering
\begin{tabular}{|p{38pt}|p{198pt}|}
\hline
\textbf{Symbol}& 
\textbf{Description}\\
\hline
            $D$             &           Mean delay\\
            $d_{max}$             &      Maximum mean delay\\
            %$E_m$           & Power consumption when using the meta-heuristic approach\\
            %$E_{on}$       & Power consumption assuming a continuous reception state\\
            $g$               & Number of epochs in the genetic algorithm\\
            $N$               &   Complexity of a certain algorithm\\
            $n_{s}, n_{l}$, $n_{sc}$             &  Number of $T_s, T_l$, and $T_{sc}$ possible values, respectively\\ 	
            %$n_{l}$             & Number of $ T_l$ possible values\\
            %$n_{sc}$             & Number of  $T_{sc}$ possible values\\
            %$\mathcal{O}(\cdot)$           & Big-O notation\\
            $P_x(t)$        &   Probability of $x$ packets arriving in $t$ units of time\\
            $p_{i,j}$       & Transition probability from a state $S_i$ to a state $S_j$\\
            %\textcolor{blue}
            {$P_q(k)$}      &   Probability of $k$ packets arriving in a burst\\
            %\textcolor{blue}
            {$p$ }        & Activation probability in the bursty traffic model\\
            %\textcolor{blue}
            {$q$ }        & Burstiness probability in the bursty traffic model\\
            ${\mathcal{S}}$       & Set of possible states\\
            $S_i$       & $i^{th}$ state in the semi-Markov chain model\\
            $S_{\text{pop}}$       & Population vector for the meta-heuristic algorithm\\
            $t$     & Unit of time\\
            %$T_{bs}$    &   Beamforming time\\  
            $T_I$       &  Inactivity time\\
            $T_l$       &  DRX long cycle duration\\
            $T_{ls}$       &  Long sleep timer\\
            $T_{on}$       &  On timer\\
            $T_s$       &  DRX short cycle duration\\
            $T_{sc}$       &  DRX short cycle timer\\
            %$T_{sh}$       &  State holding time\\
            $T_{ss}$       & Short sleep timer\\
            $U_i$	     &  Holding time for $S_i$\\
            $\beta$     & Cardinality of the Population vector\\
			$\lambda$			&	Packet arrival rate in the Poisson process\\
			$\pi_{i}$		&	Steady-state probability of being in a state $S_i$\\	
\hline
%\multicolumn{3}{p{251pt}}{Vertical lines are optional in tables. Statements that serve as captions for the entire table do not need footnote letters. }\\
%\multicolumn{3}{p{251pt}}{$^{\mathrm{a}}$Gaussian units are the same as cg emu for magnetostatics; Mx $=$ maxwell, G $=$ gauss, Oe $=$ oersted; Wb $=$ weber, V $=$ volt, s $=$ second, T $=$ tesla, m $=$ meter, A $=$ ampere, J $=$ joule, kg $=$ kilogram, H $=$ henry.}
\end{tabular}
\label{tab1}
\end{table}

{{This paper illustrates the importance of managing data buffers even in situations where DRX parameters are adjusted according to traffic patterns. To achieve this, we propose the introduction of micro sleeps that can substantially save battery life. This is particularly useful when the DRX parameter configuration lacks precision. Our approach includes the IT handling, goes beyond optimizing the DRX parameters, and is compatible with existing DRX configuration frameworks and diverse IIOT traffic patterns. The results show that by including the IT handle, the energy-saving capabilities of the DRX mechanism can be further improved, even when using the optimal DRX parameter configuration.} To the best of our knowledge, no previous research has explicitly addressed the IT reset phenomenon from the perspective of BS. Our main contributions are summarized as follows:
        \begin{itemize}
            \item We propose a novel approach that directly addresses the issue of the IT's reset from the perspective of the BS via the control channel. %This particular aspect has not been addressed by previous research. 
            %Our primary goal is to reduce 
            Our proposal reduces 
            energy consumption by %maximizing energy efficiency and 
            minimizing the time spent in a high consumption state unnecessarily.
            %\item Our primary goal of the proposed approach is to reduce energy consumption. The method attempts to maximize energy efficiency and reduce overall energy consumption, preventing the device from remaining in a high consumption state for longer than necessary.
            \item We introduce a DRX modeling based on a semi-Markov chain, while two different data traffic patterns are considered: (i) Poisson traffic and (ii) bursty traffic. This modeling methodology provides a systematic approach to studying and comprehending the behavior of the DRX mechanism in different traffic situations.
            \item %We study explicitly the effects of controlling the IT reset process from the BS and present a dynamic configuration of parameters related to DRX. 
            We %explicitly 
            investigate the effects of controlling the IT reset process from the BS and present a dynamic configuration of parameters related to DRX.  
            \item We illustrate the effectiveness of the proposed method in terms of energy savings and delay reductions. Additionally, the complexity of the proposed mechanisms is assessed, which provides a comprehensive understanding of the computational requirements.
        \end{itemize}
        Overall, the analysis and evaluation demonstrate the effectiveness of the proposed method and highlight its potential benefits in terms of energy efficiency.}
Table~\ref{tab1} lists the symbols used throughout this article. % in alphabetical order. 
The remaining part of this article is organized as follows. Section~\ref{sec2} characterizes the DRX mechanism, while Section~\ref{sec3} provides a Markovian representation. Section~\ref{sec4} presents the different IIoT traffic models and the performance metrics used in this paper. Section~\ref{sec5} formulates the optimization problem and introduces the proposed IT handling scheme. Section~\ref{sec6} describes the framework for evaluating the system performance and illustrates numerical results. Finally, conclusions are drawn in Section~\ref{sec7}. %Tables II and III list respectively acronyms and symbols used throughout this paper. For simplicity, we have omitted the well-known acronyms.  

%\begin{equation}E=mc^2.\label{eq}\end{equation}

%Be sure that the symbols in your equation have been defined before the 
%equation appears or immediately following. Italicize symbols ($T$ might refer 
%to temperature, but T is the unit tesla). Refer to ``\eqref{eq},'' not ``Eq. \eqref{eq}'' or ``equation \eqref{eq},'' except at the beginning of a sentence: ``Equation \eqref{eq}  is $\ldots$ .''

\section{DRX mechanism}\label{sec2}

\begin{figure}[t!]	% h-here, t-top, b-bottom
    \centering
    \includegraphics[width=0.95\columnwidth]{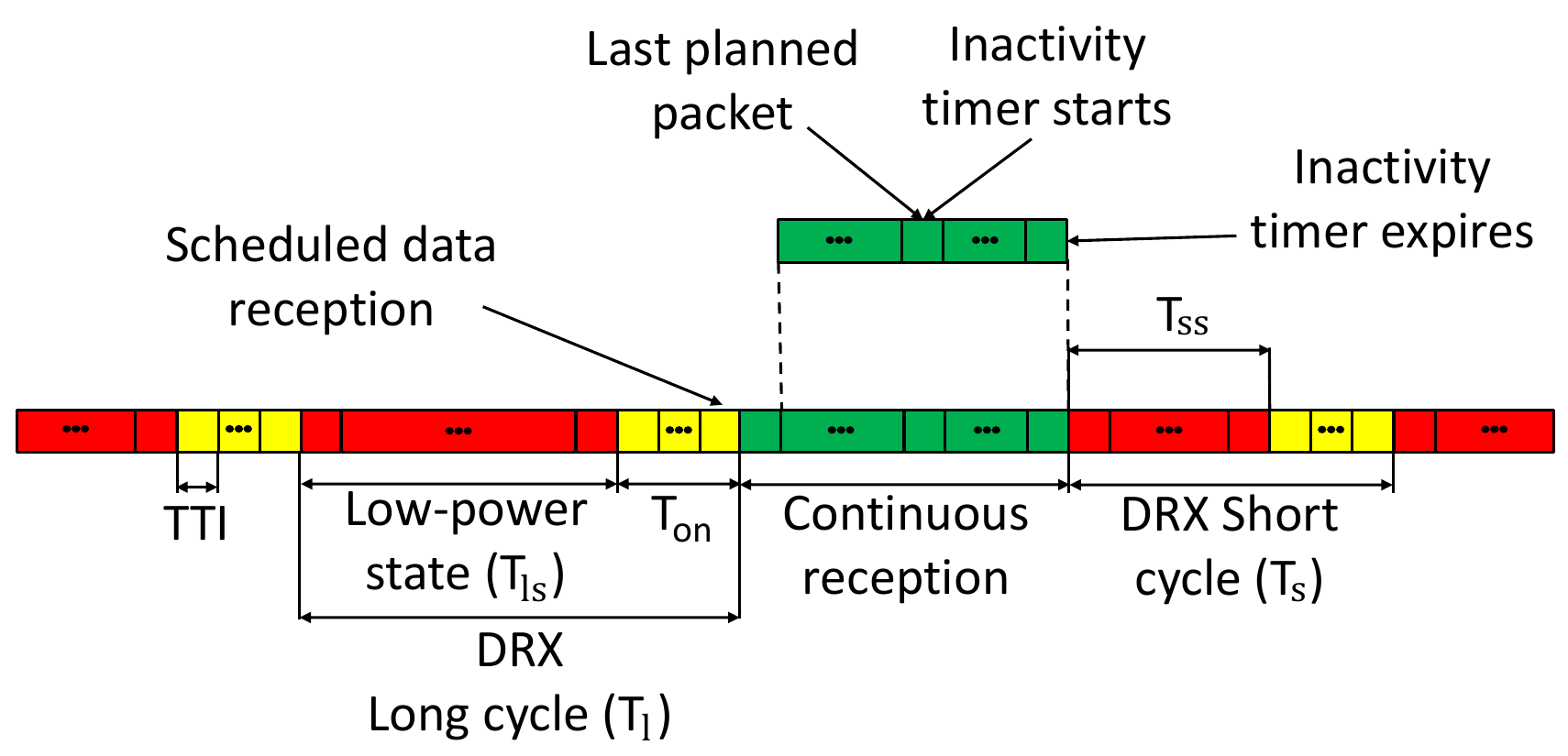}
    \caption{DRX mechanism. There are three states: (green) high consumption state in continuous reception, (yellow) high consumption state during $\text{T}_{\text{on}}$, and (red) low-power state where the interface does not read the downlink control channel. Note that the number of TTIs associated with the low-power state in the DRX long cycle is greater than that associated with that state in the DRX short cycle.}
\label{fig.1}
\end{figure}

{Fig.~\ref{fig.1} illustrates the basic DRX functioning over time slotted in transmission time intervals (TTIs). A device implementing the DRX mechanism checks the PDCCH information for a period equal to $T_{on}$~\cite{kanj2020tutorial}. If no data reception is scheduled for that time the terminal goes to a low-power state, during which the control channel is not checked~\cite{ML_DRX}. The time that the terminal remains in the low-power state is configurable. Commonly, two types of cycles of DRX operation are used, short and long cycles, characterized by their duration in the low-power state (see Fig. 1). If data reception is scheduled during $ T_{on} $,  the terminal enters the reception mode, where the control channel is continuously monitored to detect incoming data signals~\cite{kanj2020tutorial}. Notice that the battery autonomy may significantly decrease when such a reception mode is extensively prolonged. On the other hand, extending the low-power mode duration (also known as the sleep time) too much has a negative implication on the latency experienced by the data~\cite{FWuS}.}

The DRX mechanism has two modes or stages: continuous reception and DRX mode. %, the duration of the first is variable and it 
The duration of the continuous reception 
is determined %by the value of the inactivity timer ($T_I$) and 
by the behavior of the downlink traffic and by the value of the IT ($T_I$). %that arrives through the downlink channel. 
While this timer is active, the device %can be considered as 
is 
operating in continuous reception mode and there is no power saving. 
In the DRX mode, the device enters in a cycle that consists of two states: (i) a state in which the device monitors paging messages on the PDCCH, %while waits for an opportunity to enter to the other state, 
and (ii) a low-power state. If no paging messages are received during the monitoring, the device enters a DRX period during which the %device omits 
reception of the PDCCH is omitted, \textit{i.e.}, low-power state~\cite{kanj2020tutorial}.

%As part of the DRX mechanism~\cite{moradi2021improving}, the 
A terminal configured with DRX can be in one of the following states:

\begin{enumerate}
    \item {%\textcolor{blue}
    {Active state, whose duration is variable and %it is 
    determined by the value of the IT and by the behavior of the traffic that arrives through the downlink channel. %While this timer is active, the device can be considered as 
    The device in the active state is considered to be 
    operating in continuous reception mode, thus, %and therefore 
    there is no power saving effect. }}
    \item {DRX long cycle %: it is made up of a period $ T_ {on} $ and a period of sleep or low-power. If we call $ T_l $ to the duration of this cycle, the low-power time ($T_{ls}$) would be equal to $ T_l - T_ {on} $.
    of duration $T_l$, which consists of:
        \begin{itemize}
            \item an on-time $ T_ {on}$, which specifies the number of consecutive TTIs in a DRX cycle. During the validity of this timer, the device monitors the PDCCH for any associated schedule messages.
            \item a long sleep or low-power ($T_{ls}$) period.
        \end{itemize}}
        Therefore, $T_l = T_{on}+T_{ls}$.
    \item {DRX short cycle 
    of duration $T_s$ has the same composition as the DRX long cycle but with shorter sleep or low-power ($T_{ss}$) period, \textit{i.e.}, $T_{ss}<T_{ls}$.
    %: it has the same composition as the long cycle, but it is shorter. If $ T_s $ is the duration of the short cycle, the low-power time ($T_{ss}$) in this cycle can be determined as $ T_s - T_ {on} $.
        %\begin{itemize}
            %\item DRX short cycle timer ($ T_ {sc} $): specifies the number of DRX short cycles that the device should follow after the DRX IT has expired. The device is in the short cycle until the short cycle timer expires. After that, the device shifts into a long cycle DRX.
        %\end{itemize}
        }
\end{enumerate}

The IT is reset when a new transmission is received %for DL or uplink (UL) 
on the PDCCH. After this timer expires, the device %should 
enters the DRX mode. %Optionally, short DRX before long DRX. 
The device is in the short cycle until the DRX short cycle timer ($ T_ {sc}$) expires. This timer specifies the number of DRX short cycles that the device should follow after the IT has expired. 
After that, the device enters a long cycle DRX. 

%The duration of each of the previous cycles, as well as all times throughout the paper, are always a whole number of TTI. 
Time durations are expressed in terms of the number of TTIs, including the duration of each of the above cycles. 
For example, %a value of 100 assigned to the IT indicates that, 
$T_I=100$ indicates that 
100 TTIs  must elapse {from} the last transmitted/received packet to {switch} from the continuous reception mode to the DRX short cycle. 
%to move from the continuous reception period to the DRX short cycle, 100 TTI intervals must elapse after the last transmitted/received packet. 
%In 5G New Radio (NR) 
Notice that 
a flexible and adaptable frame structure has been considered in 5G New Radio (NR) to efficiently multiplex various services~\cite{won20213gpp}. The flexibility is %not only in the frequency domain with scalable orthogonal frequency division multiplexing (OFDM) numerology, but 
also in the time domain with scalable TTI~\cite{pedersen2021overview}. %It should be noted that
Therefore, 
the same numerical value assigned to any of the timers that handle the DRX mechanism can refer to different absolute values of time depending on the value of TTI\footnote{%\textcolor{blue}
{Note that once the TTI %the TTI value 
is defined, we assume that its value is uniform throughout the network for each run.}}. %In this research, TTI values of 1, 0.5, 0.25 and 0.1 ms are considered. 
%It is been shown in~\cite{eDRX} 
Note 
that reducing the idle time can significantly increase the battery lifetime and operability of the devices~\cite{eDRX}. However, this is constrained by the IT configuration, which acts as insurance against unwanted delays and packet losses. Still, in a %DL 
downlink 
scenario, a device configured with DRX could take advantage of the %knowledge provided by the base station 
BS's knowledge 
about the incoming traffic to determine whether or not is necessary to wait for the inactivity timer to expire before entering the sleep state. The exploitation of this information %in accordance with 
and 
the flexibility provided by 5G NR is a promising line of investigation~\cite{binsalem}, %, apcc}, 
which we explore next.  

\section{Markovian Representation of DRX}\label{sec3}

\begin{figure}[t!]
    \centering
    \centerline{\includegraphics[width=\columnwidth]{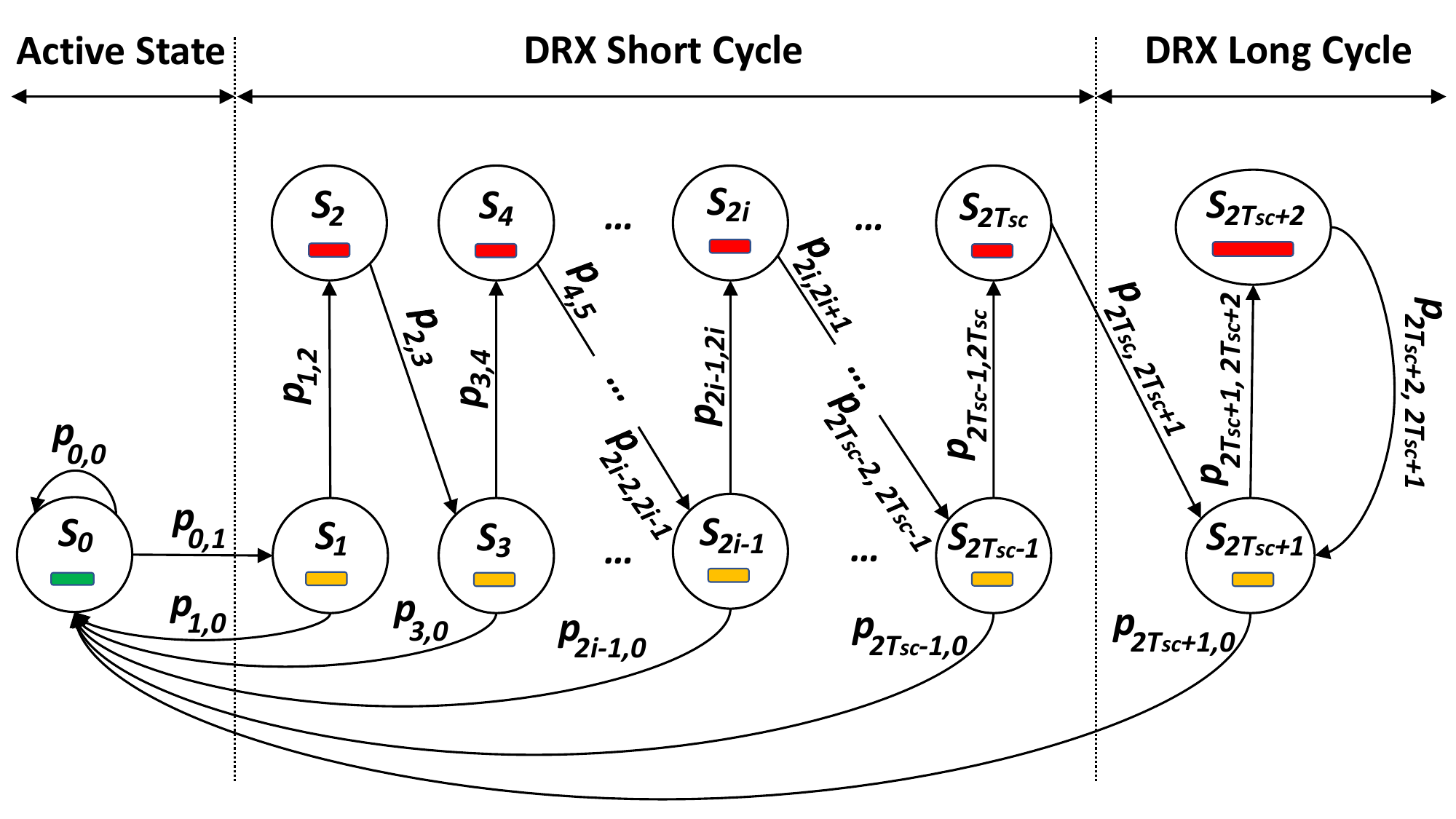}}
    \vspace{-2mm}
    \caption{%\textcolor{blue}
    {Semi-Markov chain model for DRX. The continuous reception %(green) 
    and $\text{T}_{\text{on}}$ %(yellow) 
    are 
    high-consumption %states, while 
    states and appear with green and yellow color, respectively, while the 
    low-power states are colored in red. Short and long cycles are distinguishable by the size of the colorbar (longer for long cycles).} %\textcolor{red}{aqui puse estos colores para compaginar con la fig 1, sin embargo podemos cambiar los colores en ambas figuras, tal vez de rojo el color gris.}
    }
    \label{figure_model_Markov}
    {\vspace{-4mm}}
\end{figure}

In this section, we present an analytical model for the DRX mechanism aforementioned. %\textcolor{blue}
{Consider a semi-Markov chain model~\cite{Zhou_2013} with %$2T_{sc}+3$ states ($S_k \in {\mathcal{S}}$), with 
a set $ {\mathcal{S}}$ %being the set 
of possible states %within the model and
$S_k$, 
$k \in [0,1,\dots,2T_{sc}+2]$, as shown in Fig.~\ref{figure_model_Markov}.} %State 0 (
$S_0$ stands for the active mode (or continuous reception mode) with variable duration%, states $2i-1$ ($S_{2i-1}, \ i \in [1, T_{sc}]$) represent 
. $S_{2i-1}$ with $i\in[1,T_{sc}]$ represents
the on-state in the $i^{th}$ DRX short cycle with duration $T_{on}$%, while states $2i$ ($S_{2i},\ i \in [1, T_{sc}]$) are the 
. Each state $S_{2i}$ with $i\in[1,T_{sc}]$ is 
a sleep or low-power state with duration $T_s - T_{on}$. Meanwhile, the state $S_{2T_{sc}+1}$ %(refered as $S_l$ to distinguish between short and long cycles) 
represents the on-state in a DRX long cycle with duration $T_{on}$, and $S_{2T_{sc}+2}$ %(refered as $S_{l+1}$ to distinguish between short and long cycles) 
is the sleep or low-power state in a DRX long cycle with duration $T_l - T _{on}$. We denote $p_{i,j}$ as the transition probability from a state $S_i$ into a state $S_j$ where $S_i, S_j \in  {\mathcal{S}}$.

Transitions between states can occur as follows:

\begin{enumerate}
    \item %When the device is in $ S_0 $, if there is not packet scheduled before the IT expires, the device is transferred to the $ S_1 $ state (related to $p_{0,1}$); otherwise, it resets the IT and remains in $ S_0 $, (with probability $p_{0,0}$).
    %\textcolor{blue}
    A transition from $S_0$ to $S_1$ occurs when there are no packets scheduled before the IT expires. %In case of no transition, the IT is reset.
    An appropriate IT handle prevents unnecessary IT resets that increase the power consumption.
    \item %When the device is in the state $ S_ {2i-1}, \ i \in [1, T_ {sc}] $, if there is not packet scheduled before the on duration expires, the device goes to state $ S_ {2i} $ and begins to sleep with probability $p_{2i-1,2i}$; otherwise it goes to $ S_0 $ with probability $p_{2i,0} $. We assume that synchronism is guaranteed in the delivery of the page occasion.
    A transition from $S_{2i-1}$ to $S_{2i}$ (same as from $S_{2T_ {sc}+1}$ to $S_{2T_ {sc}+2}$) occurs when there are no packets scheduled before the ON duration expires. %In case of no transition to $S_{2i}$, it goes 
    Otherwise, the transition is 
    to $S_0$. %We assume that synchronism is guaranteed. %in the delivery of the page occasion.
    \item %When the device is in the state $ S_ {2i}, \ i \in [1, T_ {sc} -1] $, after sleeping for a period $T_{ss}$, it wakes up and shifts to the state $ S_ {2i + 1 } $, related to $p_{2i,2i+1}$.
    A transition from $S_{2i}$ to $S_{2i+1}$ occurs once the sleep period $T_{ss}$ ends. 
    %The transition lasts a time $T_{bs}$ 
    %related to the beamforming time. %which here is assumed to be 1 TTI. 
    %\item %When the device is in the $ S_ {2T_ {sc}} $ state, after sleeping for a period of $ T_ {ss} $ it is transferred to the $ S_ {l} $ state to start the first DRX long cycle with probability $p_{2T{sc},2T{sc}+1}$.
    %A transition from $S _ {2T_ {sc}}$ to $S_ {2T_ {sc}+1}$ occurs once the sleep period $T_{ss}$ ends.
    %\item %When the device is in the $ S_ {l} $ state, if there is not packet scheduled before the on duration expires, it is transferred to the $ S_ {l + 1} $ state and begins to sleep with probability $p_{l,l+1}$; otherwise, the user equipment is transferred to $ S_0 $, ($p_{l,0}$).
    %A transition from $S_{2T_ {sc}+1}$ to $S_{2T_ {sc}+2}$ occurs when there are no packets scheduled before the ON duration expires. %In case of no transition to $S_{ 2T_ {sc}+2}$, it goes 
    %Otherwise, the transition is 
    %to $S_0$. 
    \item %When the user equipment is in the $ S_ {l + 1} $ state, after a sleep periodof $ T_ {ls} $ it wakes up and goes to the $ S_ {l}$ state, related to $p_{l+1,l}$.
    A transition from $S _{2T_ {sc}+2}$ to $S_{2T_ {sc}+1}$ occurs once the sleep period $T_{ls}$ ends. %The transition lasts a time $T_{bs}$. %which is assumed to be 1 TTI. 
\end{enumerate}

\noindent{Note that when the device is in a $T_{on}$ state ($S_{2i-1}$), %$\forall i \in [1, T_{sc}+1]$, 
there are three possible situations regarding the packet arrival:}

\begin{itemize}
    \item %the packet arrives after the expiration of
    %\textcolor{blue}
    {no packets arrive during 
    $T_{on}$,
    \item a packet arrives %in the $j^{th}$ subframe of
    during $T_{on}$,
    \item the packet arrived during the last sleep period. %In such case, the state hold time in ON state %if the packet arrived during the last sleep period 
    %is denoted by $T_{sh}$. 
    If a packet arrives during the sleep period of the $(i-1)^{th}$ DRX short cycle, it is delivered in the first subframe (1 TTI) of the next ON state.} %Therefore, the next ON state lasts %$ T_ {sh} $ 
    %only 1 TTI before going to state $ S_0 $.
\end{itemize}

%\subsection{Steady-state probabilities}

%\textcolor{red}{reduje el tamano de los subscript en las ecuaciones}

\section{Traffic models and Performance metrics}\label{sec4}

Machine-type communication, typical in IIoT setups, has three elementary traffic patterns~\cite{lopez2021csi}: (i) periodic update (PU), under which {devices transmit status reports %on a regular basis
regularly, \textit{e.g.}, smart meter reading (gas, electricity, water)}; (ii) event-driven (ED), which describes non-periodic traffic due to a %certain 
specific random trigger at an unknown time, \textit{e.g.}, alarms; and (iii) payload exchange (PE), which consists of {bursty traffic that usually comes after PU or ED traffic.} 
%In practice, MTC traffic often appears as a combination of the aforementioned traffic types~\cite{moh2022smart}. Hence, using the three elementary classes above for traffic modeling enables building models with an arbitrary degree of computational complexity and accuracy~\cite{eslam}. 
%\textcolor{blue}
{In this paper, we model the %transmission activation probability 
traffic 
in two different ways: (i) Poisson traffic, and (ii) %a two-state Markov chain
bursty traffic, which are illustrated in Fig.~\ref{figure_Markov_chain} and characterized below. As shown in Fig.~\ref{figure_Markov_chain}, the traffic is described as a two-state Markov chain where the idle state represents no packets arrival while the active state represents the arrival of at least one packet. {Table~\ref{traffic} summarizes the %statistical description 
key statistics 
of both traffic models.}} 

\subsection{Poisson traffic}

%We assume, without losing the generality, 
Assume 
an ED traffic pattern such that the packet arrival process follows a Poisson process\footnote{Indeed, the Poisson distribution is widely accepted for modeling the uncoordinated data traffic in heterogeneous IIoT applications~%\cite{rech2021coordinated, azari2021interference, farag2022distributed, labrador4192645raptor, ivanova2022performance}
\cite{rech2021coordinated, labrador4192645raptor, ivanova2022performance}.} %, emulating an ED traffic pattern, 
with mean $\lambda$ packets/s (pkt/s)%. Therefore, the inter-packet interval %($t$) 
%follows an exponential distribution,
, thus, 

\begin{equation}
	P_x(t) = \frac{(\lambda t)^x}{x!}  e^{-\lambda t}, \ x = 0, 1, 2, \dots
\end{equation}
 
\begin{figure}[t!]
    \centering
    \centerline{\includegraphics[width=0.75\columnwidth]{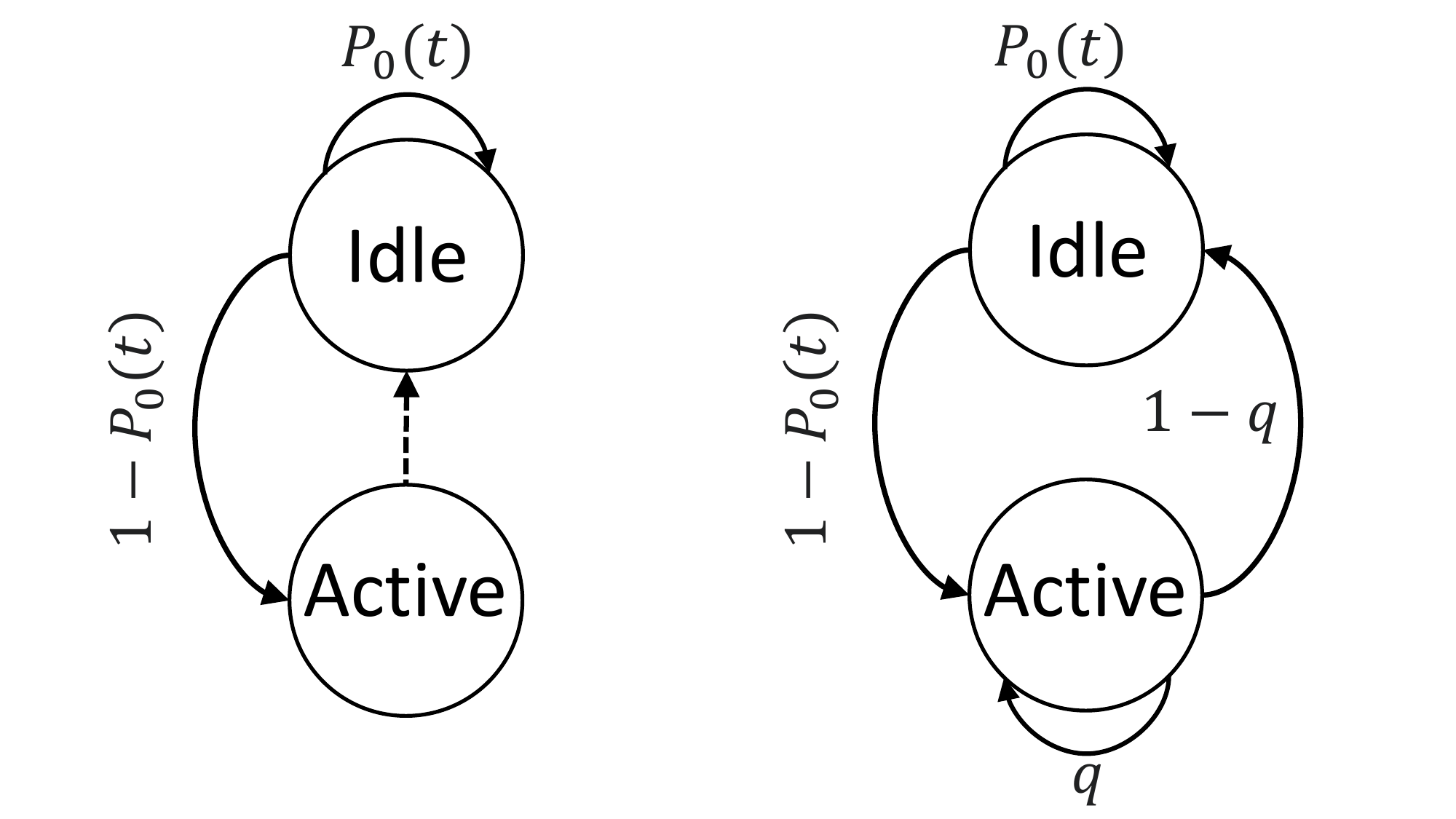}}
    \vspace{-2mm}
    \caption{a) Poisson traffic model where dashed line means %immediate return to idle state, while $p_\lambda$ represents the activation probability according to the mean rate $\lambda
    instantaneous return 
    (left), and b) %two-state Markov chain 
    bursty traffic model (right) for traffic generations. %Idle (I) and Active (A) states. 
    Every TTI in the active state means a packet arrival. %The exhaustive search finds the best combination of parameters through brute force, showing the best performance possible for each model.
    }
    \label{figure_Markov_chain}
\end{figure}

\noindent Therefore, the inter-packet interval follows an exponential distribution with expected value $1/\lambda$. %The probability of $x$ packets arriving in $t$ units of time is given by:
Notice that the probability of no packet arrival within $t$ is $ P_0(t) = e^{-\lambda t}$. %Therefore we can obtain $p_{i,j}$ as follows
%\begin{subequations}
%\begin{alignat}{2}
    %& p_{0,1} &&= p(t > T_I) = P_0(T_I) = e^{-\lambda T_I}, \\
    %& p_{0,1} &&= P_0(T_I) = e^{-\lambda T_I}, \\
    %& p_{2i-1,2i} &&= e^{-\lambda T_{on}}, \ i \in [1, T_{sc}],\\
    %& p_{2i, 2i+1} &&= e^{-\lambda (T_{ss}+T_{bS})},\ i \in [1, T_{sc}],\\
    %& p_{l,l+1} &&= e^{-\lambda T_{on}},\\
    %& p_{l+1,l} &&= e^{-\lambda (T_{lS}+T_{bS})},
%\end{alignat}
%\label{Transition_Prob}
%\end{subequations} 
%\noindent where $T_{ss} = T_s - T_{on}$ and $T_{lS} = T_l - T_{on}$. 
%Here, $T_{bs}$ is related to the beamforming time, which is assumed to be 1 TTI.  %Since $T_{ss}$ and $T_{ls} >> T_{bs}$
%Now, the holding time $U_i$ for each $S_i %\in \mathbf{\{S\}}
%%\begin{subequations}
%\begin{alignat}{2}
    %& U_{0} &&= {(1 - p(t > T_I))}\frac{1}{\lambda} = \frac{1 - e^{-\lambda T_I} }{\lambda}, \\
    %& U_{0} &&= {(1 - P_0(T_I))}\frac{1}{\lambda} = \frac{1 - e^{-\lambda T_I} }{\lambda}, \\
    %& U_{2i-1} &&= \frac{(1 - e^{-\lambda T_{on}}) }{\lambda},\ i \in [1, T_{sc}],\\
    %& U_{2i} &&= T_{ss}+T_{bs},\\
    %& U_{l} &&= \frac{1 - e^{-\lambda T_{on}} }{\lambda},\\
   % & U_{l+1} &&= T_{lS}+T_{bS}.
%\end{alignat}
%\label{Holding_time}
%\end{subequations} 
Fig.~\ref{figure_Markov_chain}a illustrates the model as a semi-Markov chain.

\subsection{{Bursty traffic}}%{Two-state Markov chain}

Assume 
an ED traffic pattern such that 
the packet arrival process is according to the Markov chain depicted in Fig.~\ref{figure_Markov_chain}b. For each TTI that the system is in the active state, a new packet is generated. Then, the probability of no packet arrival within $t$ units of time is given by 
\begin{equation}
	P_0(t) = p (1-p)^{t-1},
%\label{}	
\end{equation}
where $p$ is the activation probability. In addition, the PE (quantified here as the time spent in an active state each time the system goes to active) pattern is modeled through the $q$ parameter in the geometric distribution. This parameter $q$ tunes the burstiness of the traffic generated.
Once active, the device remains in this state for a number $k$ of TTIs (PE),  geometrically distributed with parameter $q$ and represented in {the form}
\begin{equation}
	{P_{q}} (k)= (1-q)q^{k}, \ k = 0,1,\dots.
\label{geometric}	
\end{equation}
The introduction of the additional parameter $q$ in the Markov chain allows tuning the temporal correlation of the total rate process.

%\textcolor{blue}
{The parameter $p$ for the bursty traffic model is calculated in a way that the mean arrival rate ($\lambda$) is the same as for Poisson model for fair comparison purposes. Then, 
\vspace{-2mm}
\begin{equation}
	\lambda = p + \sum_{k=1}^{\infty} p(1-q)q^k.
\label{fair}	
\end{equation}
\vspace{-4mm}
%$\lambda = p + \textstyle\sum_{k=1}^{\infty} p(1-q)q^k$. 
Also, note that 
%\begin{equation}
	$\sum_{k=1}^{\infty} q^k = q/(1-q)$,
%\label{fair2}	
%\end{equation}
%$\textstyle \sum_{k=1}^{\infty} q^k = q/(1-q)$ 
so
$\lambda = p + pq = p(1+q)$
and 
\begin{equation}
	p= \lambda/(1+q).
\label{fair3}	
\end{equation}
%$p= \lambda/(1+q).$
}

%Following the steady-state probability, we can conclude that:
%\begin{equation}
	%(1-q)\Pi_A  = p\Pi_I.
%\label{markov}	
%\end{equation}
%\noindent where $\Pi_A$ and $\Pi_I$ are the active and idle state probabilities.

%\vspace{-5mm}

\begin{table}[t]
        \caption{{Traffic Statistics Description}}
        %\label{traffic}
        \setlength{\tabcolsep}{3pt}
        \centering
        \begin{tabular}{|l|c|c|}%|p{1.4in}|p{0.6in}|p{0.6in}|}
        \hline
        \textbf{{Statistic}}& \textbf{{Poisson}}& 
        \textbf{{Bursty}}\\
        \hline
            {{Mean arrival rate}}&
            %\hline
            {$\lambda$}& {$p(1+q)$}\\
            %\multicolumn{2}{|c|}
            {{Probability of no packet arrival}}&
            %\hline
            {$e^{-\lambda t}$} &{$p (1-p)^{t-1}$}\\
            %\hline
            %\multicolumn{2}{|c|}
            {{Probability of $k$ consecutive packets arrival}}&
            %\hline
            {$\frac{(\lambda k)^k}{k!}  e^{-\lambda k}%, \ x = 0, 1, 2, \dots
            $}&      {$(1-q)q^{k}%, \ k = 0,1,2,\dots
            $}\\
            %\hline
            %\multicolumn{2}{|c|}
        \hline
        \end{tabular}
        \label{traffic}
        \end{table}
\vspace{-3mm}
\subsection{Performance metrics}\label{sec31}

Let $\pi_k$ with $k \in [0,1,\dots,2T_{sc}+2]$ be the steady-state probability of being in $S_k$ {(in the Appendix)}, %{\footnote{{See the Appendix for mathematical derivations on the steady-state probabilities.}}, and} 
and consider the probability density function (PDF) of %the inter-packet interval ($t$), which provides the probabilities of 
$x$ packets arriving in $t$ units of time is given by
%\begin{equation}
	$P_x(t)$. %, \ x = 0, 1, 2, \dots,$
%\end{equation} 
%for all possible values of $x$. 
Then, the state transition probabilities ($p_{i,j}$) are %calculated as follows.
given by
\begin{subequations}
\begin{alignat}{2}
    %& p_{0,1} &&= p(t > T_I) = P_0(T_I) = e^{-\lambda T_I}, \\
    p_{\scriptscriptstyle 0,1}& &&= P_0(T_I), \\
    p_{\scriptscriptstyle 2i-1,2i}& &&= P_0(T_{on}), \ \forall i \in [1, T_{sc}+1],\\
    p_{\scriptscriptstyle 2i, 2i+1}& &&= P_0({T_{ss}}),\ \forall i \in [1, T_{sc}],\\
    %& p_{l,l+1} &&= e^{-\lambda T_{on}},\\
    p_{\scriptscriptstyle T_{sc}+2,T_{sc}+1}& &&= P_0({T_{ls}}),
\end{alignat}
\label{Transition_Prob}
\end{subequations}where $T_{ss} = T_s - T_{on}$ and $T_{lS} = T_l - T_{on}$.  
%Here, $T_{bs}$ is related to the beamforming time, which is assumed to be 1 TTI.  %Since $T_{ss}$ and $T_{ls} >> T_{bs}$
Now, the holding time $U_i$ for each $S_i %\in \mathbf{\{S\}}
$ %are given by
is obtained as 
\begin{subequations}
\begin{alignat}{2}
    %& U_{0} &&= {(1 - p(t > T_I))}\frac{1}{\lambda} = \frac{1 - e^{-\lambda T_I} }{\lambda}, \\
    U_{\scriptscriptstyle 0}& &&= {(1 - P_{\scriptscriptstyle 0} (T_I))}/{\lambda}, \\
    U_{\scriptscriptstyle 2i-1}& &&= {(1 - P_{\scriptscriptstyle 0} (T_{on}))}/{\lambda},\ \forall i \in [1, T_{sc}+1],\\
    U_{\scriptscriptstyle 2i}& &&= T_{ss},\ \forall i \in [1, T_{sc}],\\
    %& U_{l} &&= \frac{1 - e^{-\lambda T_{on}} }{\lambda},\\
    U_{\scriptscriptstyle 2T_{sc}+2}& &&= T_{ls}.
\end{alignat}
\label{Holding_time}
\end{subequations}where $\lambda$ is the mean packet arrival rate.

\subsubsection{Power saving factor}\label{sec311}

To assess the performance of our proposal, we study the %trade-off between the 
achievable 
power saving %achieved while maintaining a set 
(PS) given a mean 
delay constraint. 
%Based on the aforementioned models, 
%we can obtain the power saving factor (PS) in the system when applying the proposed model. 
PS is calculated as the proportion of time that the device is in a low-power (sleep) state ($S_{2i},\ i \in [1, T_{sc}+1]$)%. Following the results derived above:
, \textit{i.e.}, 
\begin{equation}
	\text{PS} =  \frac{\sum\limits_{i=1}^{T_{sc}+1} \pi_{\scriptscriptstyle 2i} U_{\scriptscriptstyle 2i}%+\pi_{l+1} U_{l+1}
	}{\sum\limits_{i=0}^{2T_{sc}+2}\pi_{i} U_{i}%+\sum\limits_{i=0}^{1}\pi_{l+i} U_{l+i}
	}.
\label{PS}
\end{equation}

\subsubsection{Mean delay}\label{sec32}

Packets arriving in sleep states have to wait until the next $T_{on}$ to be sent, thus suffering a delay due to the DRX mechanism. %Since packet arrival follows a Poisson distribution
%Then, t
The mean delay of the system %($ D$) 
is given by 
\begin{align}
%\begin{array}{rl}
   D =  &\frac{ D_{T_{ss}}  \sum\limits_{i=1}^{T_{sc}}\pi_{\scriptscriptstyle 2i}U_{\scriptscriptstyle 2i} + D_{T_{ls}}\pi_{\scriptscriptstyle 2T_{sc}+2}U_{\scriptscriptstyle 2T_{sc}+2}  }{%\sum\limits_{i=0}^{2T_{sc}}\pi_{i}U_{i}+\sum\limits_{i=0}^{1}\pi_{l+i}U_{l+i}
   \sum\limits_{i=0}^{2T_{sc}+2}\pi_{i}U_{i}
   }. %\nonumber\\
	 %&+ \frac{\pi_{l+1}U_{l+1} D_{T_{ls}} }{\sum\limits_{i=0}^{2T_{sc}}\pi_{i}U_{i}+\sum\limits_{i=0}^{1}\pi_{l+i}U_{l+i}}.
%\end{array}
\label{delay_eq}
%\end{equation}
\end{align}
In the case of Poisson traffic, when a packet arrives during a sleep state in a DRX short cycle $S_{2i} \ (i \in [1, T_{sc}])$, the mean delay can be calculated as $T_{ss}/2$. Similarly, when a packet arrives during a sleep state in a DRX long cycle ($S_{l+1}$), 
the mean delay can be calculated as $T_{ls}/2$. Thus, 
$D_{T_{ss}}=T_{ss}/2$ 
and $D_{T_{ls}}=T_{ls}/2$. 
%while for the bursty traffic %it is given by
%\begin{equation}
	 %D =  \frac{T_{ss}\sum\limits_{i=1}^{T_{sc}}\pi_{2i}U_{2i}  + T_{ls} \pi_{2T_{sc}+2}U_{2T_{sc}+2}  }{2\sum\limits_{i=0}^{2T_{sc}+2}\pi_{i}U_{i} %+\sum\limits_{i=1}^{2}\pi_{2T_{sc}+i}U_{2T_{sc}+i}\right)
	 %}.
%\label{delay_P}
%\end{equation}
%Meanwhile, for two-state Markov chain traffic, the %delay is obtained as follows. %The average delay for 1 packet burst is
%\begin{equation}
    %D(1) = \frac{(1-q)q}{T_{ss}}\sum_{d_1 = 1}^{T_{ss}}d_1.
%\end{equation}
%From $d_1$, $D(2)$ is obtained as
%\begin{align}
    %D(2) &= \frac{(1-q)q^2}{T_{ss}-1}\sum_{d_1 = 2}^{T_{ss}}D(2|d_1) \nonumber\\
    %&= \frac{(1-q)q^2}{T_{ss}-1}\sum_{d_1 = 2}^{T_{ss}}\sum_{d_2 = 1}^{d_1-1} \frac{d_1+d_2}{d_1-1}.
%\end{align}
%From this derivations, 
%We can obtain the 
{Meanwhile, the mean delay %in %$T_{ss}$ ($D_{T_{ss}}$) as
%$T$ 
for the bursty traffic} 
is given by 
\begin{equation}
    D_{T} = \sum_{k = 1}^{T}\frac{(1-q)q^k}{T-(k-1)}\sum_{d_1 = k}^{T}F(k),
\end{equation}
%and the mean delay in $T_{ls}$ ($D_{T_{ls}}$) as
%\begin{equation}
    %D_{T_{ls}}(k) = \sum_{k = 1}^{T_{ls}}\frac{(1-q)q^k}{T_{ls}-(k-1)}\sum_{d_1 = k}^{T_{ls}}F(k),
%\end{equation}
where $T$ can be $T_{ss}$ or $T_{ls}$, and 
\begin{equation}
    F(k) = \sum_{d_2 = k-1}^{d_1-1} \sum_{d_3 = k-2}^{d_2-1} \dots \sum_{d_k = 1}^{d_{k-1}-1} \frac{\sum_{i=1}^{k}d_i}{\prod_{i=1}^{k-1}(d_i-(k-i))}.
\end{equation}
Here, $d_k$ is the delay of a $k^{th}$ packet within a DRX cycle. 
For example, the average delay for 1 packet burst in a DRX short cycle is
\begin{equation}
    D_{T_{ss}} = \frac{(1-q)q}{T_{ss}}\sum_{d_1 = 1}^{T_{ss}}d_1 = \frac{q(1-q)(T_{ss}+1)}{2}.
\end{equation}
%Then, substituting $T_{ss}/2$ and $T_{ls}/2$ by $D_{T_{ss}}$ and $D_{T_{ls}}$ respectively in~\eqref{delay_P} the mean delay is given by  
%\begin{equation}
    %\begin{split}
	 %D =  &\frac{ (1-q)\left(\sum\limits_{k=1}^{\infty} (\frac{T_{ss}}{2}-k)q^k \right) \sum\limits_{i=1}^{T_{sc}}\pi_{2i}U_{2i}}{\sum\limits_{i=0}^{2T_{sc}}\pi_{i}U_{i}+\sum\limits_{i=0}^{1}\pi_{l+i}U_{l+i}} \nonumber \\
	 %&+ \frac{\pi_{l+1}U_{l+1}(1-q) \left(\sum\limits_{k=1}^{\infty} (\frac{T_{ls}}{2}-k)q^k \right)}{\sum\limits_{i=0}^{2T_{sc}}\pi_{i}U_{i}+\sum\limits_{i=0}^{1}\pi_{l+i}U_{l+i}}.
	 %\end{split}
%\label{delay}
%\end{equation}
\vspace{-3mm}
\section{Configuration of DRX parameters}\label{sec5}

We assume that the terminal explicitly declares whether it %has the ability to 
can 
use the DRX mechanism during %the initial phase of exchange of 
an initial 
control information exchange phase between the terminal and the BS. If so, the BS is in charge of determining and informing the terminal, through the control channel, the configuration of the DRX parameters%that it should use in the configuration of the DRX mechanism. These parameters include the
, \textit{i.e.}, 
%$ T_ {on} $, $T_I$, 
$ T_s $, $ T_l $, IT, and $ T_ {sc}$, to be used. %The result obtained from the application of the DRX mechanism is determined by the correct configuration of these parameters.
{Our aim is to maximize PS while maintaining the delay constraints. Next, we present two methods for configuring the DRX parameters. The first method involves an exhaustive search to find the optimal configuration, leading to a lower bound for the power consumption when using the classical DRX mechanism. Meanwhile, the second method, which is contributed in this paper, exploits a less computationally expensive metaheuristic to find a sub-optimal configuration. Additionally, we propose for the BS to explicitly manage the IT reset for each device, and describe how to incorporate both the DRX parameter configuration and the IT handling. Finally, we present some practical considerations.}

\subsection{Optimization methods}

%In this paper, the parameters are calculated as follows, for each TTI value () under analysis and for a given arrival rate, the combination that allows obtaining the greatest power savings without exceeding a certain limit in terms of latency, declared as a goal, is explored using numerical methods. In this sense, these parameters can be considered optimal.

The goal is to maximize the PS without exceeding a mean delay, $d_{max}$%. Which is formulated as the following optimization problem
, thus, the problem is formulated as
%\begin{maxi}|l|
%	  {T_s, T_l, T_sc}{PS}{}{}
%	  \addConstraint{d}{\leq d_{max},}{}
%	  \addConstraint{l(w_k)}{=5u,\quad}{k=0,\ldots,N-1}
%\end{maxi}
\vspace{-1mm}
\begin{subequations}
    \begin{alignat}{2}
    & \underset{T_s,T_l,\text{IT},T_{sc}}{\text{maximize }}\:
    & &\: \text{PS} \\
    & \text{subject to }\:
    & &\: D \leq d_{max}, \\
    &&&\: T_{s} < T_l.
    \end{alignat}
    \label{opt}
\end{subequations}

%\begin{subequations}
%\begin{alignat}{2}
%	&P1: \max_{T_s, T_l, T_sc} PS\\
%    &\subjectto_{i=1}^{4} p_{i} = 1,
%\end{alignat}
%\label{optimization}
%\end{subequations}
\vspace{-3mm}
\noindent {Each parameter has impact on the system performance including $T_s, T_l, T_{sc}$, and the IT (see Section~\ref{sec31}). On the one hand, a shorter sleep time (\textit{i.e.,} $T_s, T_l$) increases responsiveness and provides lower latency, while a longer sleep time reduces energy consumption. On the other hand, a smaller $T_{sc}$ decreases energy consumption, while larger values provide lower latency by increasing cautiousness (prioritizing conservative decision-making to meet latency constraints). Similarly, shorter IT decreases the energy consumption but may increase the expected delay.}
{%Ultimately, the impact of each DRX parameter on the system performance can vary depending on the specific network conditions and quality-of-service (QoS) objectives. Therefore, 
It is essential to carefully tune and optimize these parameters based on the unique characteristics of the network deployment to achieve the desired balance between energy efficiency and latency performance.}

{Note that average latency constraints may not be enough in certain delay-sensitive applications. Instead, probabilistic guarantees on instantaneous delay realizations may be needed. Nevertheless, this may require knowledge of the delay distribution, which is extremely difficult to obtain in general. Nonetheless, we can still use our problem formulation by imposing a mean delay constraint and leveraging Markov's inequality~\cite{ghosh2002probability}. 
This allows us to establish the following theoretical relationship 
\begin{equation}
    1-F(d) \leq \frac{\mathcal{E}[d]}{d_{max}}, 
\end{equation}
where $\mathcal{E}[d]$ is the expected value of the delay (mean delay) and $F(d)$ the cumulative density function (CDF) of the instant delay. This inequality allows us to derive an upper bound on the probability of the instantaneous delay surpassing the desired mean delay threshold. This analysis provides valuable insights into the system's performance and its ability to meet latency requirements~\cite{delay_bound}. For example, when employing a mean delay constraint of 10 ms, the probability of experiencing an instantaneous delay greater than 50 ms is less than 20\%. Similarly, with a mean delay constraint of 5 ms, this probability decreases below 10\%.}

Unfortunately, solving problem~\eqref{opt} is not straightforward since the power consumption and delay are not jointly convex on the optimization parameters. 
Herein, we resort to two numerical methods for solving~\eqref{opt}. 

\subsubsection{Brute Force (exhaustive search)}

The brute force method requires testing every combination in which $ T_s $, $ T_l $, and $ T_ {sc} $ can be drawn. Thus, the complexity %($N$) 
is 

\begin{equation}\label{eq1}
    N = \mathcal{O}\left(n_{s} n_{l} n_{{sc}}\right),
\end{equation}

\noindent{being $n_s, n_{l}$, and $n_{sc}$ the number of possible values for $ T_s $, $ T_l $, and $ T_ {sc} $, respectively~\cite{ML_DRX}}. 
Therefore, %when the brute force holds, we are able to attain 
the optimum DRX configuration is obtained at the end of the longest-possible exploration. This includes multiple independent simulations of arrival generation and transmission or service time for each combination of parameters. Notice that it might be very computationally expensive to run an exhaustive search.

\subsubsection{Sub-optimal optimization}

%Since the brute force method takes the longest-possible exploration, 
Herein
we %aim to find a more suitable approximation with less computational cost. 
present a sub-optimal solution for \eqref{opt}, which is more computationally affordable than the brute force method. 
We %choose the %MATLAB 
%meta-heuristic algorithm tools (Genetic Algorithm (GA)) %and Particle Swarm Optimization (PSO))
specifically resort to a generic algorithm (GA), 
which relies on evolutionary mechanisms of natural selection~\cite{banerjee2022impacts}.
{GAs are recognized for their effectiveness in exploring large solution spaces and handling nonlinear non-convex problems. Moreover, GAs are parallelizable, scalable, and robust to noise or uncertainty in objective functions~\cite{du2022multiuser} and adapt to changing environments and dynamic problems~\cite{golberg1989genetic}. In fact, the GA model selection and optimization process depend less on specific variables or factors, allowing for greater flexibility and adaptability while maintaining transparency, thus significantly different from back-box approaches~\cite{zhao2022adaptive,allahloh2023optimizing}. By utilizing a GA, we can address the limitations of standard convex optimization techniques, which may struggle with highly non-linear, non-convex, and dynamic problems~\cite{allahloh2023iiot}.}
The pseudo-code is illustrated in Algorithm~\ref{alg1}. %shows the pseudo-code follow by the GA algorithm tool.

Such a kind of meta-heuristic algorithms are %are an approach to optimization and learning 
based loosely on principles of biological evolution, are simple to construct, and their implementation does not require a large amount of storage. %, making them a sufficient choice for an optimization problems. 
%Meta-heuristic algorithms have been shown 
They are known 
to solve linear and nonlinear problems efficiently by exploring all regions of the state space and exponentially exploiting %promising areas through 
the application of mutation, crossover, and selection operations to individuals in the population~\cite{saraswat2013genetic}.

By using the meta-heuristic algorithm approximation, we obtain a near-optimum configuration with 

\begin{equation}
    N = \mathcal{O}(g \beta),
\end{equation}

\noindent{where $g$ is the maximum number of epochs/iterations and $\beta$ is the number of elements (cardinality) of the candidate population vector (group of DRX parameter values, $S_{\text{pop}}$).  %which is set up to 200, 
{%In each generation, the fitness of every parameter is evaluated according to %the power consumption and mean delay using an specific configuration. 
%how well a particular parameter set performs in terms of power consumption and mean delay, \textit{i.e.,} lowest power consumption while the mean delay is below the delay constraint.
In each generation, the fitness of every parameter is evaluated based on how well a specific parameter set performs in terms of power consumption and mean delay. Specifically, the evaluation aims to achieve the lowest power consumption while ensuring that the mean delay remains below the delay constraint. } 
The best fit parameters are stochastically selected from the current population, and each parameter is modified (recombined and possibly randomly mutated) to form a new generation (new combination of DRX parameters). The new generation of candidate solutions is then used in the next iteration of the algorithm~\cite{saraswat2013genetic}. Commonly, the algorithm terminates when either a maximum number of generations ($g$) has been produced, or a satisfactory fitness level has been reached for the population.
%Therefore, the algorithm can stop before reach this limits if an optimum value is obtained.
}
\textcolor{black}{\begin{algorithm}[t]
\caption{Genetic algorithm pseudo-code}\label{alg1}
\begin{algorithmic}
\State \textbf{Generate} an initial population of individuals ($S_{pop}$, combination of DRX parameters)
\State \textbf{Evaluate} the fitness of all individuals {(parameter sets with the lowest power consumption)}
    \While{termination condition not met} %\& iteration $< g$} %\And{iteration < g} %{best individual $= \varnothing$}
        \State{\textbf{Select} fitter individuals for reproduction (select DRX parameters configuration with less power consumption and mean delay below the constraint)
        \State \textbf{Recombine} individuals (form new combination of DRX parameters with the selected values in the previous step)}
        %\State \textbf{Mutate} individuals
        \State \textbf{Evaluate} the fitness of the modified individuals (power consumption and mean delay)
        \State \textbf{Generate} a new population by using the remaining selected values in previous steps and new values of DRX parameters different to the ones used as previous individuals
    \EndWhile
\end{algorithmic}
\end{algorithm}}

Notice that despite the high cost in computational use, %notice that 
the optimum configuration parameters can be tuned for each traffic pattern and TTI value. Therefore, the optimization phase can be developed offline and the values stored at the BS. Then, when the system is online, a %look at a table 
lookup table 
(previously configured in the training process) is only needed for setting the DRX configuration with the optimum or at least near-optimum values. Thus,
%\begin{equation}
    $N = \mathcal{O}(1)$,
%\end{equation} 
\noindent which is the only computational cost needed online\footnote{Note that the memory cost of keeping and updated table is %infimum and not worthy to take into account.
{negligible.}}. However, notice that even though the offline optimization phase can be developed using any of the previous methods (exhaustive search or meta-heuristic), %notice that in case of retraining due to changes in traffic patterns,
the second option (meta-heuristic) is much less computationally costly in the case of retraining due to changes in the traffic patterns while the system is online. 

%\end {itemize}

\subsection{Inactivity Timer control}

\begin{figure}[t]
    \centering
    \includegraphics[width=0.75\linewidth]{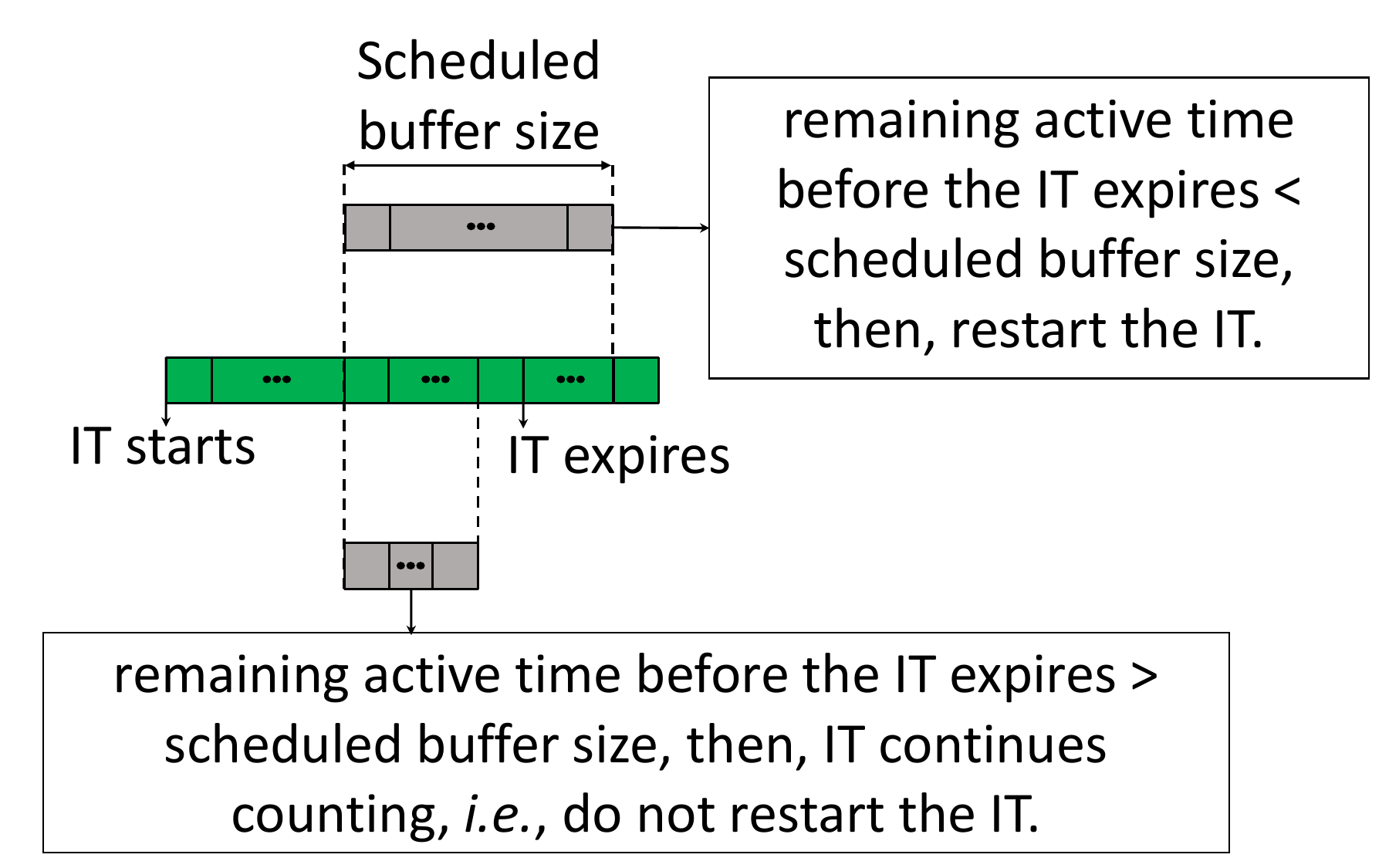}
    \vspace{-1mm}
\caption{Proposed inactivity timer handling linked to transmitter (Tx) buffer state (per-TTI scheduling buffer at the Tx).}
\label{buffer}
\end{figure}

The traditional DRX mechanism works autonomously in the terminal itself. {However, in this research, we propose that the BS explicitly declares to each terminal when an IT reset is required through the control channel. We exploit the fact that the BS knows the volume of information contained in its buffer and the transmission rate assigned to the terminal. This allows the BS to foresee whether it is possible to transmit all the pending scheduled information in the time remaining until the IT expires. If infeasible, the BS indicates the reset of the IT, otherwise the IT reset is not requested to avoid unnecessary PDCCH-only slots without any grant. However, any packet that arrives at the BS during the period of low-power consumption of the target device must wait in the buffer until a new active period. It is worth noting that this modification is binary and has negligible additional weight on the control channel.}
The situations linked to the buffer's state and the IT handle are illustrated in Fig.~\ref{buffer}.

{The proposed method differs from the classical DRX as the BS dynamically indicates whether the IT should be reset during each data transmission in the downlink channel based on real-time conditions. The proposal is supported by the knowledge of the data buffer state and the downlink scheduling, which enables the calculation of the volume of data to be transmitted.} Note that the proposed method does not require a precise characterization of the traffic handled between the BS and the devices, but only statistical. From now on, we refer to the mechanism that incorporates the aforementioned features as `\textit{intelligent IT handling}'  while we refer to the mechanism that performs a standard IT reset as indicated in the standard DRX mechanism as `\textit{standard IT}'. {The pseudo-code for our proposed method is illustrated in Algorithm~\ref{alg2}}.

\begin{algorithm}[h]
\caption{Proposal pseudo-code}\label{alg2}
{\begin{algorithmic}
%\vspace{-2mm}
\State \textbf{Estimate} traffic statistics and 
\textbf{Configure} DRX parameters through optimization methods
    \While{online} 
        \If{BS has information to transmit \& device active} 
            \State {BS \textbf{exchange information} with the  device}
            \If{remaining time $>$ scheduled buffer size}
                \State {BS indicates not to restart the IT}
            \EndIf
            \State \textbf{Check} traffic statistics and \textbf{Re-configure} DRX
            \State 
            parameters  \textbf{if} traffic statistics change
        \EndIf
    \EndWhile
\end{algorithmic}}
\end{algorithm}

\subsection{{Practical considerations}} 

{We assume that the traffic statistics (\textit{i.e.,} $\lambda$, $p$, and $q$) are perfectly known by the BS. This assumption allows us to optimize the DRX parameters effectively in addition to the intelligent IT handling. Indeed, the BS may monitor the buffer's status to gather information about the traffic load and accurately estimate the traffic parameter statistics. These estimations are crucial for efficiently planning and managing network resources, ensuring optimal performance and a satisfactory user experience. However, the estimation process is separate from the IT handling, which relies solely on the knowledge of pending data (buffer's status) for each device to determine the resource allocation.}
        
{For simplicity, we focus on the performance analysis of the proposed method for a single-user scenario, aimed at increasing the understanding of the problem and the proposed mechanism. However, the approach can be extended straightforwardly to multiple users by leveraging the BS ability to differentiate between packets destined for different users and manage the IT on a per-user basis, which however increases the complexity. As the number of users increases, both the GA-based and exhaustive search-based strategies experience a corresponding increase in complexity when configuring the DRX parameters. This is assuming that each device has independent traffic, \textit{i.e.,} if there are \textit{M} devices, the complexity becomes $M\times N$. However, this complexity is unrelated to the intelligent IT handling, which remains straightforward even with multiple users.}
%\textchange{Algorithm }
\begin{table}[t!]
\caption{Simulation parameters}
\label{table1}
\begingroup \centering
%\processtable{Parameters used in the simulation.}
    \begin{tabular}{|l|l|l|}
    \hline
    \textbf{Parameter}                       & \textbf{Value}     &\textbf{Ref.}\\
    \hline
    $d_{max}$    
    & 10 (ms)     &\cite{smartNIC}\\
    $g$     
    &200    &\cite{saraswat2013genetic,lotf2022improved}\\
    %\textcolor{blue}{$p$%~\footnotemark[4]
    %&0.0025, 0.005, 0.01, 0.025         &\eqref{fair3}}\\
    %\textcolor{blue}
    $q$
    &0.5  &\cite{FWuS}\\
    $S_{\text{pop}}$       
    &50     &\cite{lotf2022improved}\\
    %packets' size       &10 Kbytes     &\cite{huawei20193gpp}\\
    %$T_{bs}$        &1 (TTI) &  \\
    $T_I$       
    & [1,1/$\lambda$] (TTI)     	&\cite{huawei20193gpp}\\
    $T_l- T_{on}$   
    &[$T_s- T_{on}$, 640] (TTI)	    &\cite{apcc,ML_DRX}\\ 
    $T_{on}$        
    & 8 (TTI)        &\cite{huawei20193gpp}\\
    $T_{sc}$        
    & [1, 16]	    &\cite{ruiz2021drx}\\ 
    $T_s- T_{on}$ 
    & [32, 160] (TTI)	&\cite{apcc,ML_DRX}\\ 
    TTI   	            & 1, 0.5, 0.25, 0.125 (ms)     	&\cite{tti,ML_DRX}\\
    $\lambda$ 
    & 5, 10, 20, 50 (pkt/s)	    &\cite{huawei20193gpp}\\ 
    \hline
    \end{tabular}{}\\
    \endgroup
    %\vspace{-1ex} 
    \
    %{\footnotemark[4]\footnotesize{    
    %\tablefootnote
    %{\textcolor{blue}{The parameter $p$ for the bursty traffic model is calculated in a way that the mean arrival rate is the same as for Poisson model for fair comparison purposes. Then, $\lambda = p + \textstyle\sum_{k=1}^{\infty} p(1-q)q^k$.}}}}
\label{table_results}
\vspace{-1ex}
\end{table}
 
\section{Performance evaluation and discussions}\label{sec6}

Numerical results are drawn and discussed in this section. Moreover, we compare the performance of the proposed and the conventional approaches. {In this paper, we focus on the advantage of handling IT to reduce power consumption, hence, comparisons with earlier works focused on DRX configuration are not pertinent, as the IT reset handle is oblivious to the DRX configuration. Instead, we aim to analyze the saving capabilities of IT handling.} 

We determine the optimal parameters of the DRX mechanism assuming that the arrival traffic to the BS follows {either} the Poisson distribution or the %two-state Markov chain 
bursty traffic 
model. 
%The maximum delay has been set at 10 ms, %and $ T_ {on} $ has a value of 8 TTI, 
The values for the short ($T_{ss}$) and long ($T_{ls}$) cycle sleep time vary between 1 and 160 TTIs in the first case, and between $T_{ss}$ and 640 TTIs in the second (the long cycle sleep value always starts at the declared short cycle sleep time value). Note that the long cycle sleep time can not be less than the short cycle sleep time by definition. %The value of $ T_ {sc} $ is varied between 1 and 16, and the 
The value of ($T_{I}$) is chosen in such a way that its duration is equal to the average time between packets ($ 1 / \lambda $) and $p$ is calculated according to~\eqref{fair3}. A constant download rate %in the downlink channel is considered, 
equal to four times the mean arrival rate is considered in the downlink channel. %A size of 
%Packets %' size 
%arriving at the BS destined for the terminal are considered to be of 0.1 Mbytes.
Table~\ref{table1} summarizes the parameters used in simulations unless explicitly stated otherwise. %It should be noted that, under such considerations, similar results would be obtained regardless of the size of the packets to be transmitted.

\vspace{-2mm}
\subsection{Power consumption analysis}

{The optimal values of the DRX parameters are determined for different arrival rates and TTIs by applying the two optimization methods outlined in Section V.A. Firstly, we utilize traffic patterns statistics to exhaustively analyze DRX parameter sets combinations to solve the optimization problem in~\eqref{opt}. Note that this is infeasible in practice because of its computational complexity but provides the best-case performance scenario for any existing studies on DRX configuration. Secondly, we employ a more computationally efficient metaheuristic approach to expedite the optimization process. In Table~\ref{CompareTable}, a brief comparison with existing studies is presented. The power consumption in this table is analyzed for TTI $=1$ ms and 10 ms delay constraint to ensure fairness. 
Performing a computationally complex Gaussian process is required in~\cite{ruiz2021drx}, while~\cite{aghdam2021traffic} requires a precise characterization of the traffic between the BS and devices.   
It is worth noting that the methods proposed in previous research are not even able to achieve the performance  obtained using the brute force method and standard IT, which is used as a comparison in this paper. This evinces the effectiveness of the metaheuristic method with IT handle even when the best method of configuring DRX parameters is the one used.} We also calculate the theoretical lower bound, which comes from assuming a genie-aided scenario, in which the device is just active when there is a transmission.

\begin{table}[t]
        \caption{{Comparison of the state-of-the-art DRX configuration for TTI $=1$ ms and 10 ms delay constraint}}
        \vspace{-1mm}
        %\label{CompareTable}
        \setlength{\tabcolsep}{3pt}
        \centering
        {\begin{tabular}
        {|p{0.58in}|p{0.73in}|p{0.38in}|p{0.32in}|c|c|}%{|c|p{0.8in}|p{0.28in}|p{0.25in}|p{0.33in}|p{0.56in}|}
        \hline
        & \centering{\textbf{{DRX Configuration}}}& 
        \centering{\textbf{{Delay-sensitive}}}& 
        \centering{\textbf{{IT handle}}}& 
        \textbf{{PS}}&
        \textbf{{Complexity}}\\
        \hline
            \centering{\cite{ruiz2021drx}} & 
            \centering{Optimized parameters with fixed IT} &
            \centering{$\checkmark$} & 
            \centering{-} & 
            {13\%}& 
            $\mathcal{O}(n^3)^{\text{\Cross}}$\\
            \centering{\cite{aghdam2021traffic}}& 
            \centering{$T_l$ and $T_s$ are variable, $T_{sc}$ and IT are fixed} & 
            \centering{-} & 
            \centering{-} & 
            {9\%}& 
            {$\mathcal{O}(n)^{\text{\Cross}}$}\\
            \centering{%\rowcolor{cyan!40}
            {{Brute Force with standard IT}}} &
            \centering{Optimized parameters} & 
            %{10ms}&
            \centering{$\checkmark$}&
            \centering{-}& 
            {21\%}& 
            $\mathcal{O}\left(n_{s} n_{l} n_{{sc}}\right)$\\
            \centering{%\rowcolor{orange}
            {{Metaheuristic with IT handling}}} &%{$\checkmark$}&
            \centering{Optimized parameters} & 
            %{10ms}&
            \centering{$\checkmark$}&
            \centering{$\checkmark$}& 
            {43\%}& 
            $\mathcal{O}(g \beta)$\\
        \hline
        \end{tabular}}
        \vspace*{1mm}
        
        \footnotesize{{\Cross `$n$' represents the dataset size for reaching convergence during training.}}
        \vspace{-4mm}
        \label{CompareTable}
\end{table}

\begin{figure*}[t]
    \centering
    \begin{subfigure}%[h]
	\centering
    \includegraphics[width=0.5\linewidth]{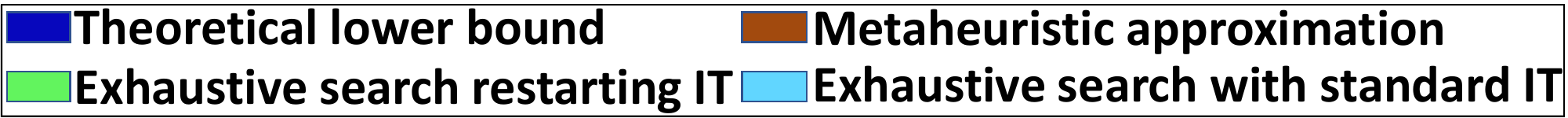}
    %\caption{Análisis de Ahorro de Potencia para $\lambda$ = 5 pkt/s.}
    \label{fig.21}
    \end{subfigure}  
    \begin{subfigure}
    \centering
    \includegraphics[width=0.49\linewidth]{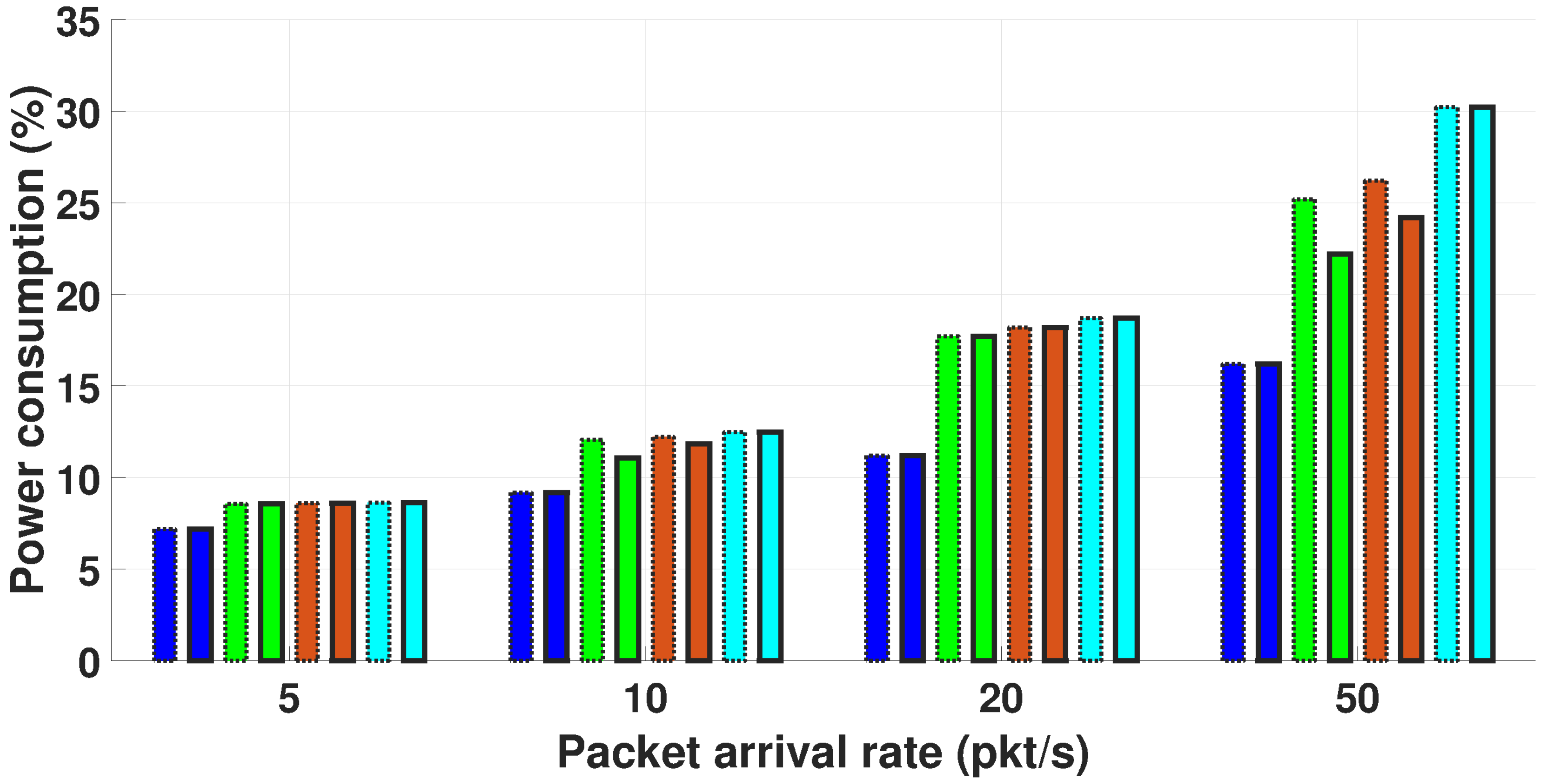}
    %\caption{Análisis de Ahorro de Potencia para $\lambda$ = 10 pkt/s.}
    \label{fig.3}
    %\vspace{-1mm}
    \end{subfigure}
    \begin{subfigure}
	\centering
	\includegraphics[width=0.49\linewidth]{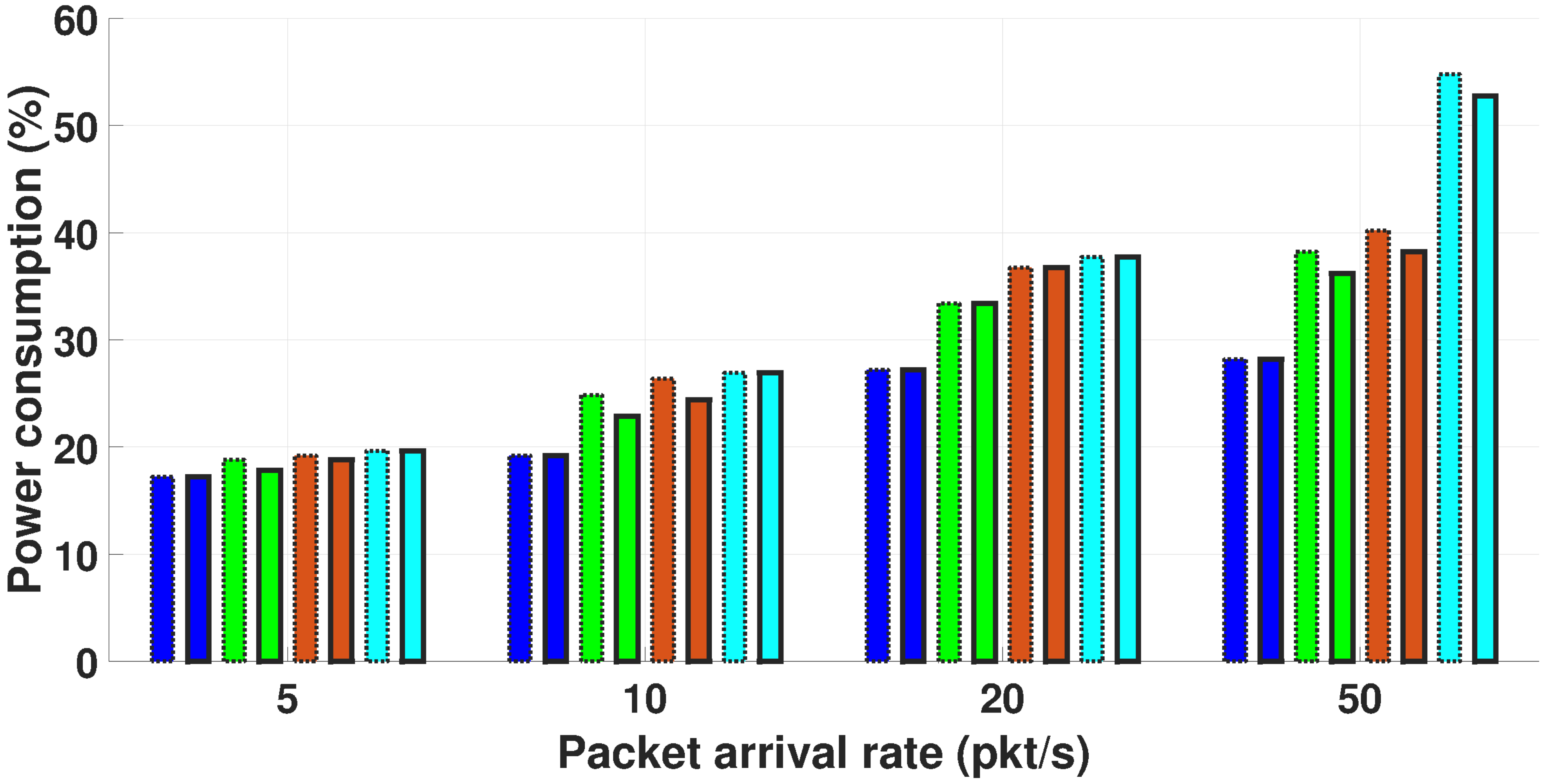}
    %\caption{Análisis de Ahorro de Potencia para $\lambda$ = 20 pkt/s.}
    \label{fig.4}
    %\vspace{-1mm}
    \end{subfigure} 
    \begin{subfigure}
	\centering
	\includegraphics[width=0.49\linewidth]{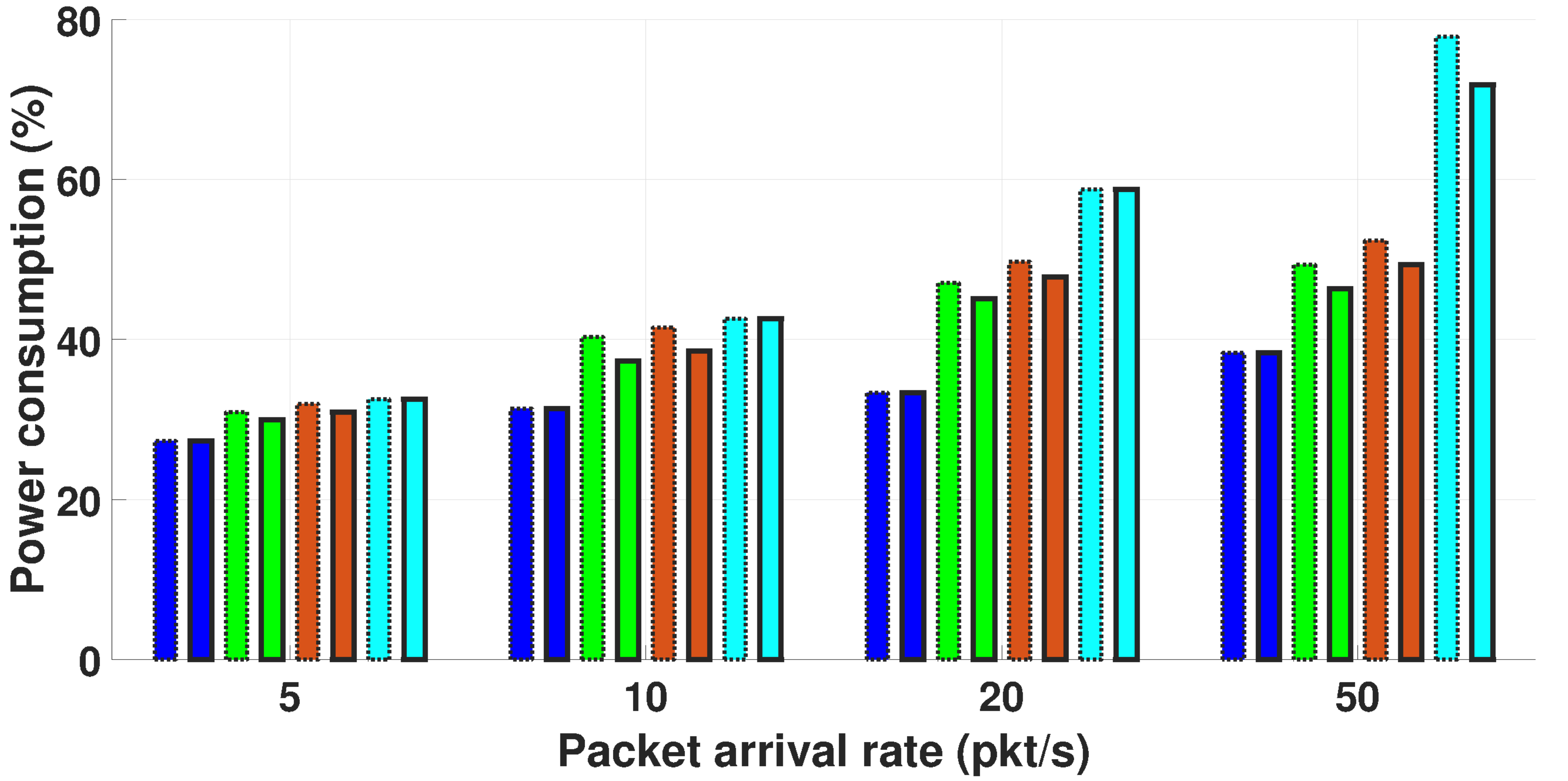}
    %\caption{Análisis de Ahorro de Potencia para $\lambda$ = 50 pkt/s.}
    \label{fig.5}
    \end{subfigure} 
    \begin{subfigure}
	\centering
	\includegraphics[width=0.49\linewidth]{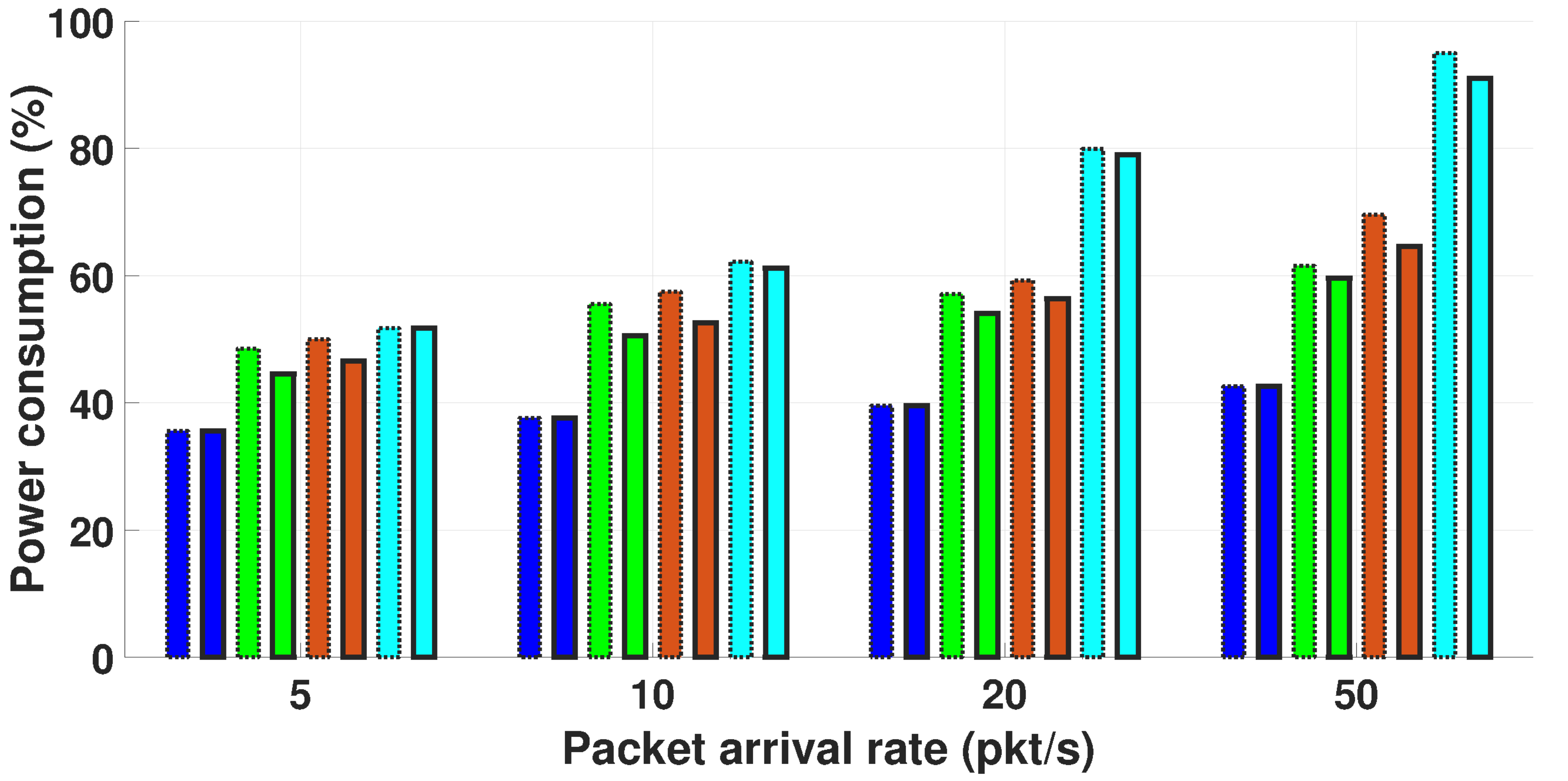}
    %\caption{Análisis de Ahorro de Potencia para $\lambda$ = 50 pkt/s.}
    \label{fig.6}
    \end{subfigure}
    \vspace{-6mm}
\caption{{Performance analysis of power consumption for }%$\lambda$ = 5 pkt/s (top left), $\lambda$ = 10 pkt/s (top right), $\lambda$ = 20 pkt/s (bottom left), $\lambda$ = 50 pkt/s (bottom right). 
Poisson traffic (bars with dotted line edges) and %two-state Markov chain 
bursty traffic model (bars with straight line edges) when the TTI is equal to a) 0.125 ms (top left), b) 0.25 ms (top right), c) 0.5 ms (bottom left), and d) 1 ms (bottom right). {The traffic rate is adjusted within the range of 5 to 50 pkt/s.}}
\label{fig.2}
\end{figure*}
%\vspace{-3mm}

For each value of $ \lambda $ (pkt/s), we compare the power consumption values obtained from the DRX mechanism with the configuration of the previous parameters following the standard restart of the IT, and those that result from applying the solution proposed in this paper. %Firstly, 
The results are presented for the Poisson arrival traffic model %(represented by the bars up in Fig.~\ref{fig.2}) while later we present the results 
and for the %Two-state Markov chain 
bursty traffic 
model.  
%taking into account 
%mimicking 
%bursty traffic. %(represented by the bars down in Fig.~\ref{fig.2}). 
The illustrated
curves are the result of Monte Carlo simulations over 250 runs, where the traffic is differently generated in each run.
Fig.~\ref{fig.2} shows the comparison of the power consumption results for  {different TTI and traffic rate values ($\lambda$) when the mean delay constraint is 10~ms. It can be seen that the power consumption increases with the TTI for the same $\lambda$, while the saving ratio between both solutions remains relatively constant.} 
It should be noted that the proposed solution always achieves power consumption values lower than the IT standard restart, reaching approximately a power consumption absolute reduction of 12$-$15\% %in absolute levels 
and of approximately 30\% relative to the traditional scheme. %It is important to recognize that 
%Nevertheless, the proposed solution implies a slight increase in latency, but this does not exceed in any case the constraint concerning the established latency. 

\begin{figure}[t]
    \centering
    \includegraphics[width=\columnwidth]{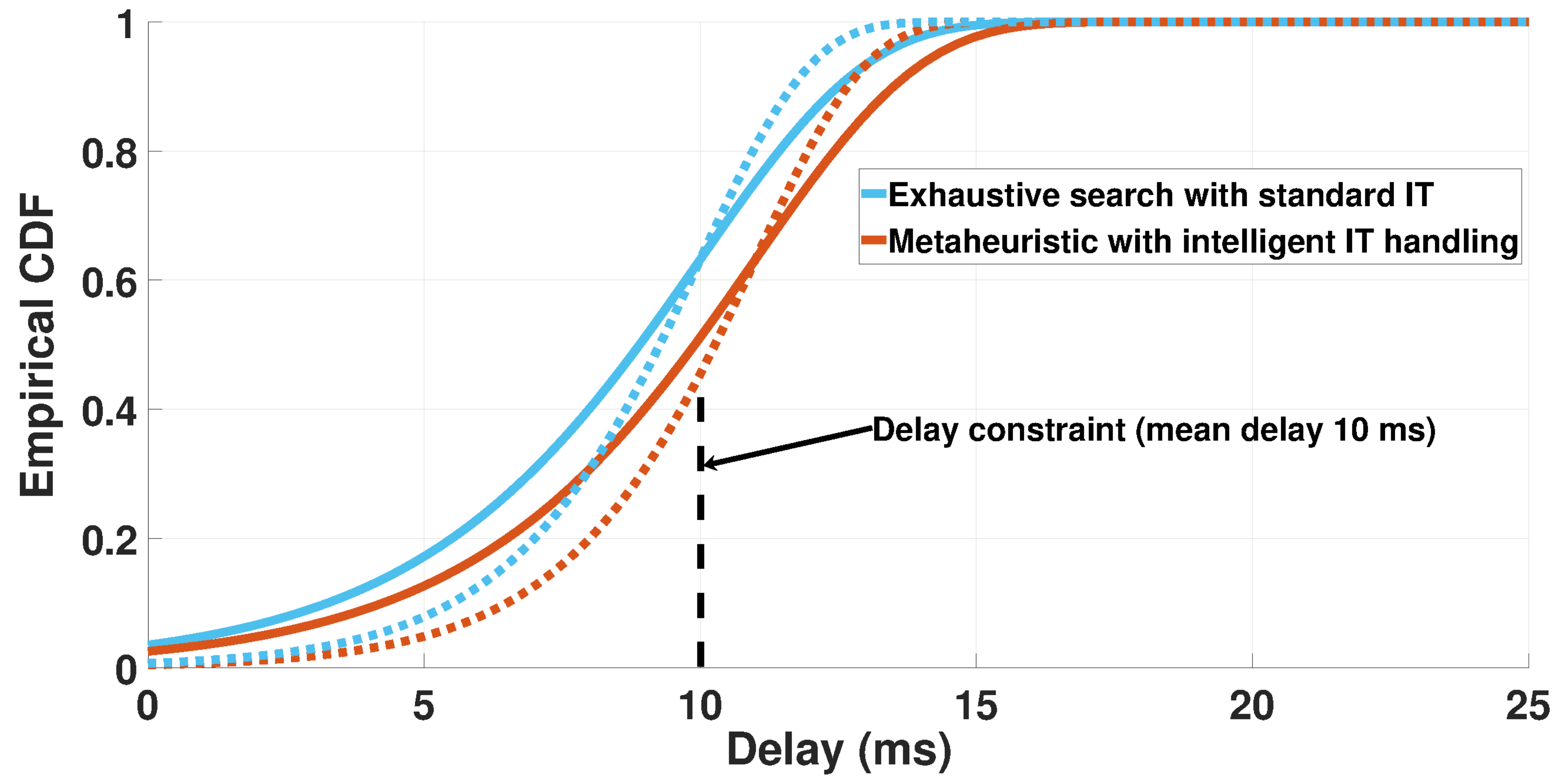}
    \vspace{-5mm}
    \caption{%Cumulative density function (CDF) 
    {CDF of the instant delay with standard IT restart and the metaheuristic approximation} for Poisson traffic (straight lines) and %two-state Markov chain 
    bursty traffic 
    model (dotted lines).}
\label{delay_cdf}
\end{figure}

It can be seen from Fig.~\ref{fig.2} that the proposed DRX configuration solution reaches values near the theoretical lower bound\footnote{{Ideal scenario for which the BS knows exactly the traffic and correspondingly handle the IT.}}. In the case of Poisson traffic and a brute force and GA-based optimization, the power consumption is reduced around 10\% and 8\%, respectively. Nevertheless, the latter optimization approach allows significantly 
%(around just 10\% more power consumption), which is calculated with our proposed scheme and the knowledge of each packet arrival instance as well as the brute force for DRX configuration. Notice that our second proposal for DRX parameters configuration, by using meta-heuristic algorithms (200 iterations of GA), attains near-optimum parameters %with much less computational cost, 
reducing the learning (training) phase. 
It is noteworthy that %(bars down), 
the results shown %in Fig.~\ref{fig.2} %similar performance for the theoretical lower bound and for the standard IT approach, while 
for the proposed restarting IT the power consumption performance decreases between 2 and 8\% when using the %Two-state Markov chain 
bursty traffic 
model. {Although the proposed solution implies a slight increase in latency, it does not exceed in any case the mean latency constraint.} 
Fig.~\ref{delay_cdf} illustrates the CDF of the instant delay where the constraint for the mean delay is marked. Notice that the instant delay increase when using the proposed mechanism does not exceed 2 ms in any case. Moreover, the instant delay is below the delay constraint more than the 50\% of the time while the maximum instant delay is below 15 ms more than 98\% of the time.

%As for the comparisons regarding the power-saving behavior of the proposed model with meta-heuristic optimization algorithms and the exhaustive search method, %while the standard IT restart model with exhaustive search optimization is also presented, providing the worst ($\lambda$ = 5 pkt/s) and best-case ($\lambda$ = 50 pkt/s) comparison for our proposal. 
All in all, 
the solution proposed in this research %(meta-heuristic) 
surpasses the conventional one, {as illustrated before in Table~\ref{CompareTable}}, regardless of the traffic rate and the TTI used. Moreover, the meta-heuristic-based DRX configuration has a similar performance to the exhaustive search method with much less computational cost. %($<1\%$). 
%For exhaustive search, 
%\textcolor{blue}
{Specifically, %$n_{l}$ depends on $n_s$, then 
$n_l n_s = \textstyle\sum_{i=32}^{160} (640-i) = 70176$ while $n_{sc} =16$, therefore 
$N = \mathcal{O}(70176 \times 16) = \mathcal{O}(1122816)$ for an exhaustive search according to~\eqref{eq1} and the values in Table~\ref{table_results}}. Meanwhile, %while for the meta-heuristic approach 
$N = \mathcal{O}(200\times50) =  \mathcal{O}(10000)$ for the meta-heuristic optimization in the worst case, which is about a 0.89\% of the previous cost. 
Notice that the gap between our proposal and the standard IT approach increases as the traffic rate and the TTI increase. It is also noteworthy that the restarting IT technique performs better when bursty traffic is involved since the standard IT method resets the IT each time a packet arrives at the device.

\vspace{-3mm}
\subsection{Delay variation} 

\begin{figure}[t!]
    \centering
    %\begin{subfigure}%[h]
	%\centering
    \includegraphics[width=\columnwidth]{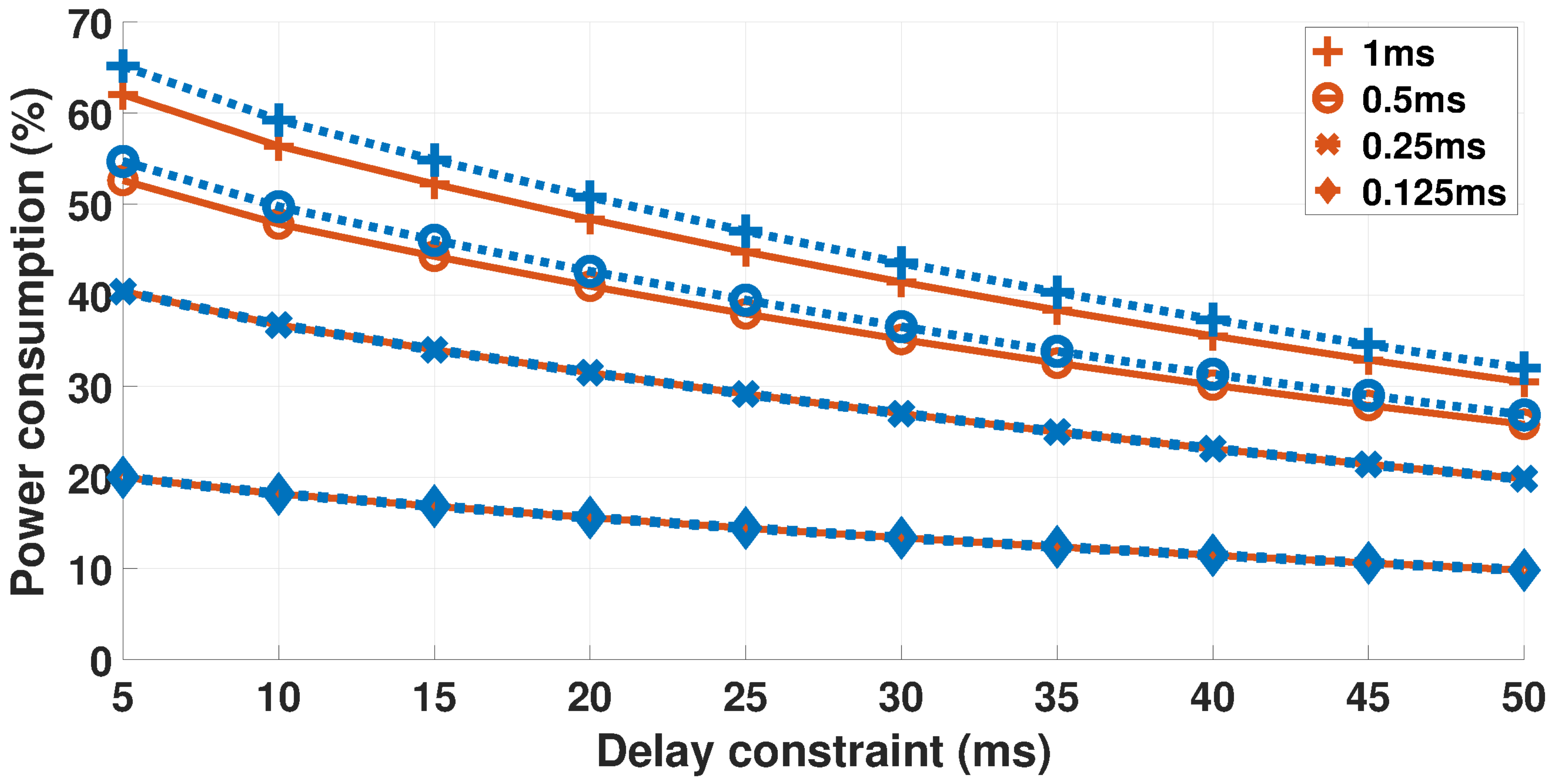}
    %\caption{Análisis de Ahorro de Potencia para $\lambda$ = 5 pkt/s.}
    %\label{fig.2}	    
    %\end{subfigure}  
    %\begin{subfigure}
    %\centering
    %\includegraphics[width=\columnwidth]{fig/delay_Markov.pdf}
    %\caption{Análisis de Ahorro de Potencia para $\lambda$ = 10 pkt/s.}
    %\label{fig.3}
    %\end{subfigure}
    %\begin{subfigure}
	%\centering
	%\includegraphics[width=0.49\textwidth]{fig/GA_5_2.pdf}
    %\caption{Análisis de Ahorro de Potencia para $\lambda$ = 20 pkt/s.}
    %\label{fig.4}
    %\end{subfigure} 
    %\begin{subfigure}
	%\centering
	%\includegraphics[width=0.49\textwidth]{fig/GA_50_2.pdf}
    %\caption{Análisis de Ahorro de Potencia para $\lambda$ = 50 pkt/s.}
    %\label{fig.5}
    %\end{subfigure}
    \vspace{-5mm}
    \caption{Relative power consumption between {meta-heuristic proposal %and exhaustive search 
    with standard restarting inactivity timer} under delay threshold variations. Under Poisson (blue dotted lines) and %Two-state Markov chain 
    bursty 
    (orange lines) traffic model, $\lambda$ = 20 pkt/s (right).}
\label{delay}
\end{figure}

%{It is important to consider delay sensitive/tolerant applications when delay requirement is small/large, and discuss how the proposed method can effectively meet these requirements~\cite{rahmani2020energy}.} %Going beyond analyzed energy-saving capabilities, Fig. 6 shows that further power saving could be achieved as the delay threshold increases, as in case of delay-tolerant applications.
%{In Fig.~\ref{delay}, it is demonstrated that increasing the delay threshold can result in additional power savings, particularly for delay-tolerant applications, surpassing the previously analyzed energy-saving capabilities. The BS assesses the performance degradation with the service requirement (delay), and can decide how to manage the configuration parameters. It is noteworthy that for delay-sensitive applications, the energy saving achieved by managing the IT (intelligent IT handling) exceeds 30\%. This means that even though the IT management mechanism by the BS needs to be more careful in handling to ensure better responsiveness and lower delay (due to the high delay constraint violation probability as shown in Fig.~\ref{delay}), the proposed method is still capable of achieving considerable savings compared to the best possible case of DRX parameters configuration with standard IT reset (Brute force). In addition,} when using a smaller TTI, less power consumption ($P_c$) is expected (around half $P_c$ for a 50 ms delay constraint). Notice that the power saving gain is even greater when the traffic follows the bursty traffic model, as it was remarked in Fig.~5.

{Small/large delay requirements correspond to delay-sensitive/tolerant applications. Herein, we discuss how the proposed method can effectively meet these requirements~\cite{rahmani2020energy}.} {In Fig.~\ref{delay}, it is demonstrated that increasing the delay threshold can result in additional power savings, especially for delay-tolerant applications, surpassing the previously analyzed energy-saving capabilities.} {The BS assesses the performance} {degradation with the service requirement (delay), and can decide how to manage the configuration parameters.} {Note that for values below 10~ms, the energy saving achieved by intelligent IT handling exceeds 30\%. This means that even though the IT management mechanism at the BS needs to ensure better responsiveness and lower delay (it may be the case for delay-sensitive applications where the system has to process and fulfill requests quickly), the proposed method achieves considerable savings compared to the best possible case of DRX parameters configuration with standard IT reset (Brute force). In addition}, when using a smaller TTI, less power consumption ($P_c$) is expected (around half for a 50~ms delay constraint). Notice that the power saving gain is even greater when the traffic follows the bursty traffic model, as it was remarked in Fig.~\ref{fig.2}.

%{By applying the simple form of Markov's inequality, we can theoretically establish a relationship between the probability of a random variable exceeding a certain value (1 - F(x)) and the ratio of the expected value of the variable ($\mathcal{E}[X]$) to that value (x)~\cite{ghosh2002probability}. This inequality allows us to derive an upper bound on the probability of the instantaneous delay surpassing the desired mean delay threshold. This analysis provides valuable insights into the system's performance and its ability to meet latency requirements. For example, when employing a mean delay constraint of 10 ms, the probability of experiencing an instantaneous delay greater than 50 ms is less than 20\%. Similarly, with a mean delay constraint of 5 ms, this probability decreases below 10\%.}
\vspace{-2mm}
\section{Conclusions}\label{sec7} 

In this paper, we showed that properly restarting the IT %, explicitly from the BS %in order 
%to reduce the active time in the DRX cycle 
has a notable impact on power savings. %while maintaining latency levels in an acceptable range. 
Specifically, we proposed a method for maximizing the energy saving by dynamically
configuring the DRX parameters. 
The %improvement is the result of 
{method} set the optimal DRX parameters and then exploits  
the knowledge that the BS has %both 
about the contents of the buffer pending to be transmitted. %and the data rate that results from the planning of the terminal. 
Notice that signaling the restart of the inactivity timer from the BS implies only a slight increase in the control information to be transmitted as such information is binary in nature, therefore it can be considered negligible. Interestingly, our proposal allows savings of up to approximately 30\% %, with respect to the standard inactivity timer restart mechanism, 
are obtained %by applying the proposal presented in this work.
with respect to the standard IT restart mechanism.

Future research is expected to extend the analysis of the IT handle by including different traffic behaviors and variable data transmission rates through the downlink channel while also considering the conditions of the channel. {In future work, we may consider the challenges of a realistic system with multiple users, incorporating factors like radio resource allocation, achievable bit rates, and adaptive DRX to accommodate varying requirements across different users. Moreover, it might be appealing to incorporate several considerations to enhance the performance and user experience for delay-sensitive applications. These include incorporating packet prioritization to ensure QoS guarantees and utilizing feedback mechanisms to improve responsiveness.}

\vspace{-3mm}
\bibliographystyle{IEEEtran}
\bibliography{BIB.bib}

\vspace{-4mm}
\appendix \label{app}

Let $\pi_k$ with $k \in [0,1,\dots,2T_{sc}+2]$ be the steady-state probability of being in $S_k$, then
\begin{subequations}
\begin{alignat}{2}
\pi_{\scriptscriptstyle 0}& &&= \frac{1}{1 - p_{\scriptscriptstyle 0,0}}\sum\limits_{i=1}^{T_{sc}+1} \pi_{\scriptscriptstyle 2i-1} p_{\scriptscriptstyle2i-1,0},\\
\pi_{\scriptscriptstyle k}& &&= \pi_{\scriptscriptstyle k-1}  p_{\scriptscriptstyle k-1,k}  %\text{ for } 
\text{ }\ \forall k \in [1, 2T_{sc}],\\
%\\
%\pi_{2T_{sc}} \cdot p_{2T_{sc},2T_{sc}+1}\\ 
%+ \pi_{2T_{sc}+2} \cdot p_{2T_{sc}+2,2T_{sc}+1}, & \text{for } i = 2T_{sc}+1 \\
%\\
%\pi_l \cdot p_{2T_{sc}+1,2T_{sc}+2}, & \text{for } i = 2T_{sc}+2 
	%& {\pi_0 = \frac{\sum\limits_{i=1}^{T_{sc}+1} (\pi_{2i-1}\cdot p_{2i-1,0})}{(1 - p_{0,0})}}\\
    %& \pi_i = \pi_{i-1} \cdot p_{i-1,i}, \ i \in [1, 2T_{sc}]\\
    \pi_{\scriptscriptstyle 2T_{sc}\!+\!1}& &&= \pi_{\scriptscriptstyle 2T_{sc}}  p_{\scriptscriptstyle 2T_{sc},2T_{sc}\!+\!1} \!+\! \pi_{\scriptscriptstyle 2T_{sc}\!+\!2}  p_{\scriptscriptstyle 2T_{sc}\!+\!2,2T_{sc}\!+\!1},\\
    \pi_{\scriptscriptstyle 2T_{sc}+2}& &&= \pi_{\scriptscriptstyle 2T_{sc}+1}  p_{\scriptscriptstyle 2T_{sc}+1,2T_{sc}+2}\\
    \sum\limits_{i=0}^{2T_{sc}+2} \pi_i& &&= 1.
\end{alignat}
\label{Steady_state_Prob}
\end{subequations}	
Then, \eqref{Steady_state_Prob}b-\eqref{Steady_state_Prob}d can be re-written as
\begin{subequations}
%$$\pi_i =
%\begin{cases}
%\pi_{0} \prod_{j=1}^{i}p_{i-1,i}, & \text{for } i \in [1, 2T_{sc}]\\
%\\
%\pi_{2T_{sc}} \cdot p_{2T_{sc},2T_{sc}+1}\\ 
%+ \pi_{2T_{sc}+2} \cdot p_{2T_{sc}+2,2T_{sc}+1}, & \text{for } i = 2T_{sc}+1 \\
%\\
%\pi_l \cdot p_{2T_{sc}+1,2T_{sc}+2}, & \text{for } i = 2T_{sc}+2 
%\end{cases}$$
\begin{alignat}{2}
    \pi_{\scriptscriptstyle k}& &&= \pi_{\scriptscriptstyle 0} \prod_{j=1}^{k}p_{\scriptscriptstyle j-1,j} \text{     }\ \forall k \in [1, 2T_{sc}],\\
    \pi_{\scriptscriptstyle 2T_{sc}+1}& &&=  \frac{\pi_{\scriptscriptstyle 0} p_{\scriptscriptstyle 2T_{sc},{2T_{sc}+1}} \prod_{j=1}^{2T_{sc}}p_{\scriptscriptstyle j-1,j}}{1 \!-\! p_{\scriptscriptstyle {2T_{sc}+1},2T_{sc}+2}  p_{\scriptscriptstyle {2T_{sc}+2},{2T_{sc}+1}}} ,\\
    \pi_{\scriptscriptstyle {2T_{sc}+2}}& &&= \frac{\pi_{\scriptscriptstyle 0} p_{\scriptscriptstyle 2T_{sc},{2T_{sc}+1}}  p_{\scriptscriptstyle {2T_{sc}+1},{2T_{sc}+2}}\prod_{j=1}^{2T_{sc}}p_{\scriptscriptstyle j-1,j}}{1 \!-\! p_{\scriptscriptstyle {2T_{sc}+1},{2T_{sc}+2}}  p_{\scriptscriptstyle {2T_{sc}+2},{2T_{sc}+1}}}  .
\end{alignat}
\label{Steady_state_Prob2}
\end{subequations}
\vspace{-5mm}
\begin{IEEEbiography}[{\includegraphics[width=1in,height=1.25in,clip,keepaspectratio]{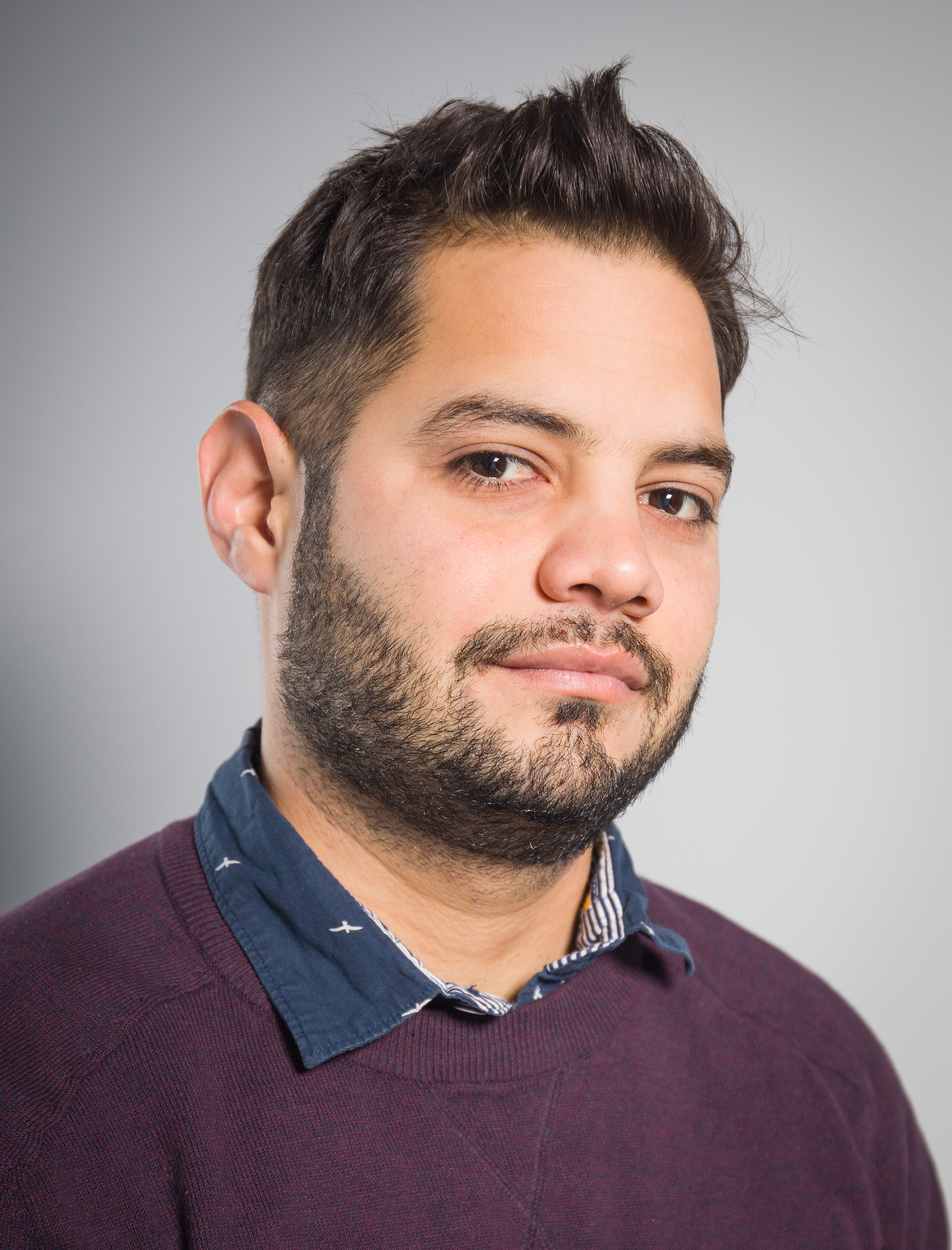}}]{David E. Ruíz-Guirola} (S'22) received the B.Sc. (1st class honors, 2018) and M.Sc. (with distinction, 2019) degree in Telecommunications and Electronic Engineering from the Central University of Las Villas (UCLV), Cuba. He is currently pursuing his Ph.D. degree at the University of Oulu. His research interests include sustainable IoT, energy harvesting, RF energy transfer, machine-type communications, machine learning, and traffic prediction.
\end{IEEEbiography}
\vspace{-5mm}
\begin{IEEEbiography}[{\includegraphics[width=1in,height=1.15in,clip,keepaspectratio]{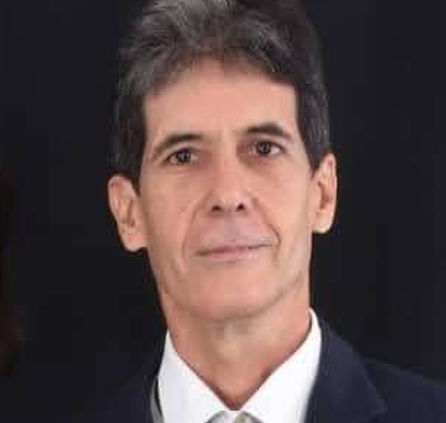}}]{Carlos A. Rodríguez-López} received the B.Sc. and M.Sc. degrees in Electronic Equipment and Components and Telecommunications from the Central University of Las Villas (UCLV), Cuba, in 1990 and 2000, respectively. He is currently a Professor with the Department of Telecommunications, UCLV. His research interests are in the area of wireless mobile communications, including visible light communications. 
\end{IEEEbiography}
\begin{IEEEbiography}[{\includegraphics[width=1in,height=1.25in,clip,keepaspectratio]{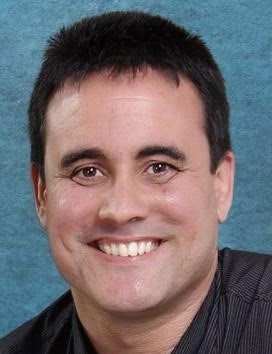}}]{Samuel Montejo-Sánchez} (M'17-SM'22) received the B.Sc., M.Sc., and D.Sc. degrees in telecommunications from the Central University of Las Villas (UCLV), Cuba, in 2003, 2007 and 2013, respectively. Associate Professor at UCLV (2003-2017). Since 2018, he has been with the Programa Institucional de Fomento a la I+D+i (PIDi), Universidad Tecnológica Metropolitana (UTEM). He leads the FONDECYT Iniciación No. 11200659 (Toward High Performance Wireless Connectivity for IoT and Beyond-5G Networks) project. His research interests include wireless communications, signal processing, sustainable IoT, and wireless RF energy transfer. He was a co-recipient of the 2016 Research Award from the Cuban Academy of Sciences.
\end{IEEEbiography}
\begin{IEEEbiography}[{\includegraphics[width=1in,height=1.25in,clip,keepaspectratio]{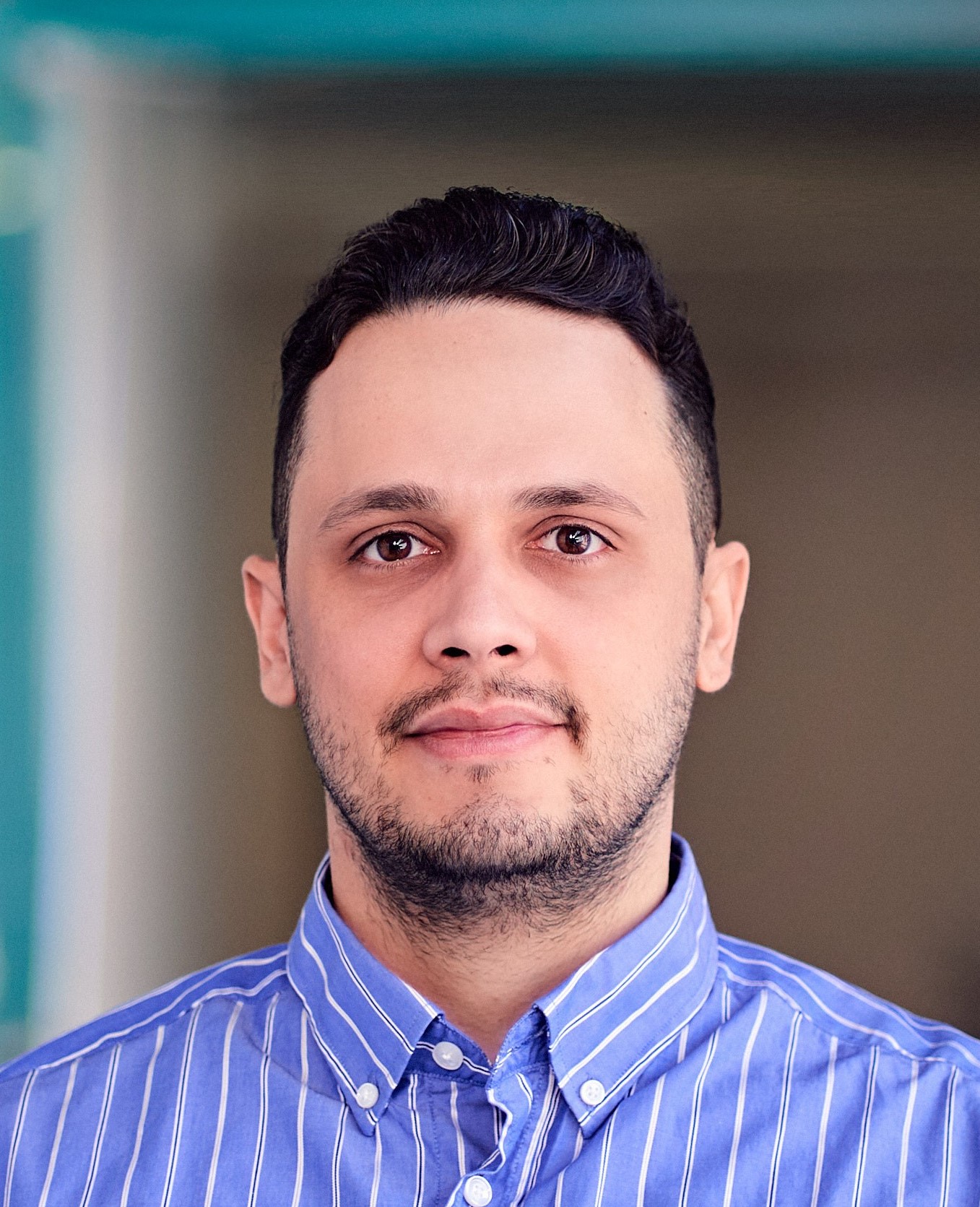}}]{Onel L. A. López} (S'17-M'20-SM'24) received the B.Sc. (1st class honors, 2013), M.Sc. (2017), and D.Sc. (with distinction, 2020) degree in Electrical Engineering from the Central University of Las Villas (Cuba), the Federal University of Paraná (Brazil), and the University of Oulu (Finland), respectively. He is a collaborator to the 2016 Research Award given by the Cuban Academy of Sciences, a co-recipient of the 2019 and 2023 IEEE European Conference on Networks and Communications (EuCNC) Best Student Paper Award, and the recipient of both the 2020 best doctoral thesis award granted by Academic Engineers and Architects in Finland TEK and Tekniska Föreningen i Finland TFiF in 2021 and the 2022 Young Researcher Award in the field of technology in Finland. Co-author of the books entitled ``Wireless RF Energy Transfer in the massive IoT era: towards sustainable zero-energy networks'', Wiley, 2021, and ``Ultra-Reliable Low-Latency Communications: Foundations, Enablers, System Design, and Evolution Towards 6G'', Now Publishers, 2023. Since 2024, he has been an Associate Editor of the IEEE Transactions on Communications. He is currently an Associate Professor in sustainable wireless communications engineering at the Centre for Wireless Communications (CWC), Oulu, Finland. His research interests include sustainable IoT, energy harvesting, wireless RF energy transfer, wireless connectivity, machine-type communications, and cellular-enabled positioning systems.
\end{IEEEbiography}
\begin{IEEEbiography}[{\includegraphics[width=1in,height=1.25in,clip,keepaspectratio]{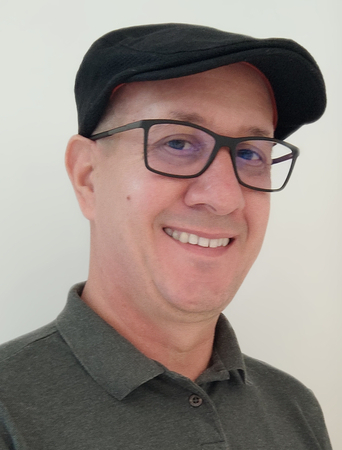}}]{Vitalio Alfonso Reguera} (Senior Member, IEEE) received the B.Sc. degree in Telecommunications and Electronics Engineering from the University of Marta Abreu of Las Villas (UCLV), Cuba, in 1995. He earned the M.Sc. degree in Telecommunications in 2000 and the Ph.D. degree in Telecommunications in 2007, both from UCLV. Full Professor in the Department of Electronics and Telecommunications at UCLV (2007 to 2018). From 2018 to 2022, he served as a Visiting Professor in the Graduate Program in Electrical Engineering at the Federal University of Santa Maria (UFSM), Brazil. He is currently a Professor in the Department of Information Technologies at the Universidad Tecnológica del Uruguay (UTEC). His research interests encompass wireless communications, IoT, and the integration of AI in advanced communication systems. He was a co-recipient of the 2016 Research Award from the Cuban Academy of Sciences. 
\end{IEEEbiography}
\begin{IEEEbiography}[{\includegraphics[width=1in,height=1.25in,clip,keepaspectratio]{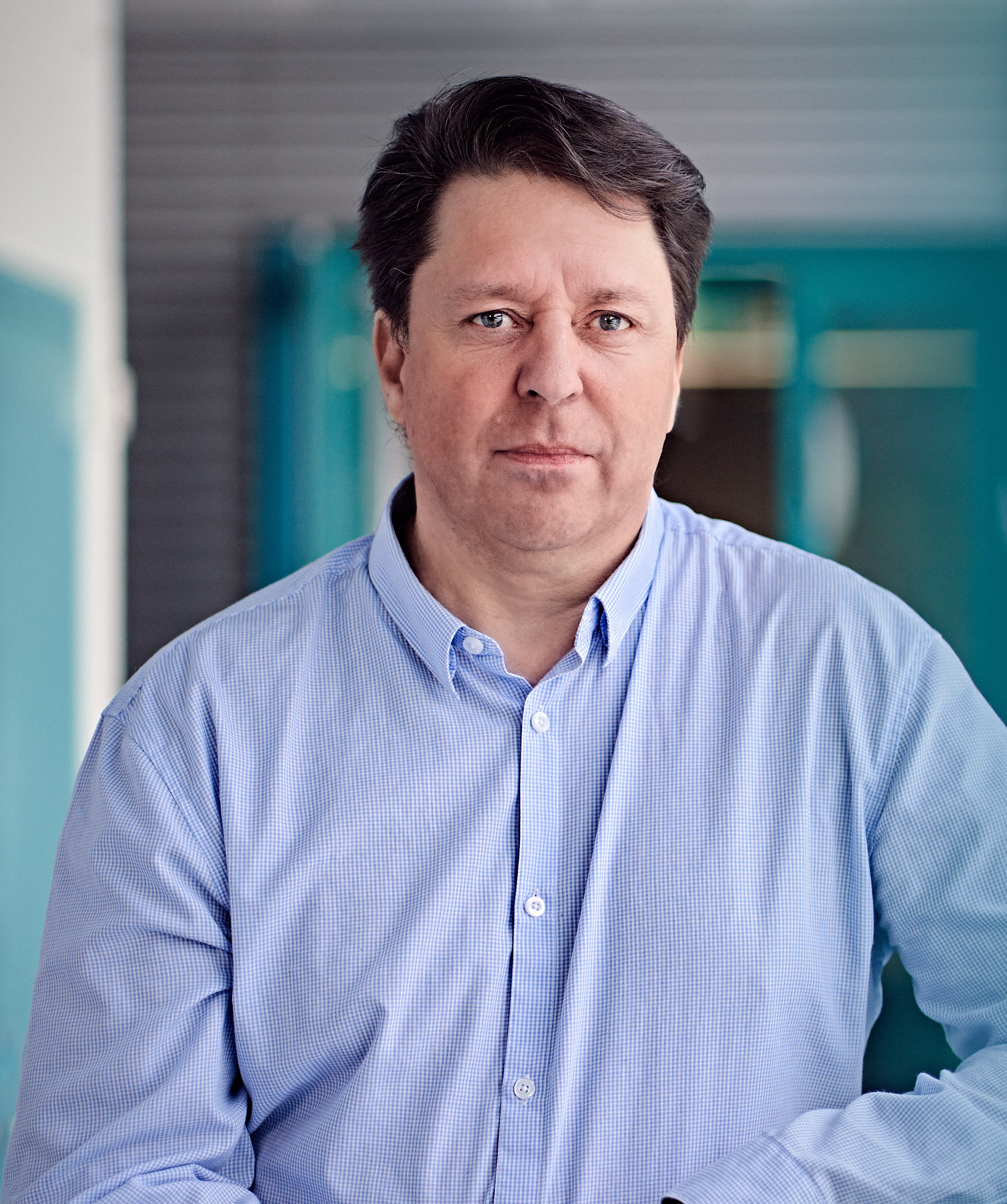}}]{Matti Latva-aho} (IEEE Fellow) received his M.Sc., Lic.Tech., and Dr.Tech. (Hons.) degrees in Electrical Engineering from the University of Oulu, Finland, in 1992, 1996, and 1998, respectively. From 1992 to 1993, he was a Research Engineer at Nokia Mobile Phones in Oulu, Finland, after which he joined the Centre for Wireless Communications (CWC) at the University of Oulu. Prof. Latva-aho served as Director of CWC from 1998 to 2006 and was Head of the Department of Communication Engineering until August 2014. He is currently a Professor of Wireless Communications at the University of Oulu and the Director of the National 6G Flagship Programme. He is also a Global Fellow at The University of Tokyo. Prof. Latva-aho has published over 500 conference and journal papers in the field of wireless communications. In 2015, he received the Nokia Foundation Award for his achievements in mobile communications research.
\end{IEEEbiography}

\end{document}